\documentclass[aps,prd,twocolumn,superscriptaddress,showpacs]{revtex4-1}
\usepackage{graphicx,amsmath,amssymb,bm,braket,afterpage,bbold}

\newcommand{\fmiq}{\, \text{fm}^{-3}}
\newcommand{\mev}{\, \text{MeV}}

\newcommand{\gevi}{\, \text{GeV}^{-1}}

\begin{document}

\title{Large-scale nuclear structure calculations for spin-dependent WIMP
scattering \\ with chiral effective field theory currents}

\author{P.\ Klos}
\email[E-mail:~]{pklos@theorie.ikp.physik.tu-darmstadt.de}
\affiliation{Institut f\"ur Kernphysik, 
Technische Universit\"at Darmstadt, 
64289 Darmstadt, Germany}
\affiliation{ExtreMe Matter Institute EMMI, 
GSI Helmholtzzentrum f\"ur Schwerionenforschung GmbH, 
64291 Darmstadt, Germany}
\author{J.\ Men\'{e}ndez}
\email[E-mail:~]{javier.menendez@physik.tu-darmstadt.de}
\affiliation{Institut f\"ur Kernphysik, 
Technische Universit\"at Darmstadt, 
64289 Darmstadt, Germany}
\affiliation{ExtreMe Matter Institute EMMI, 
GSI Helmholtzzentrum f\"ur Schwerionenforschung GmbH, 
64291 Darmstadt, Germany}
\author{D.\ Gazit}
\email[E-mail:~]{doron.gazit@mail.huji.ac.il}
\affiliation{Racah Institute of Physics and
The Hebrew University Center for Nanoscience and Nanotechnology, 
The Hebrew University, 91904 Jerusalem, Israel}
\author{A.\ Schwenk}
\email[E-mail:~]{schwenk@physik.tu-darmstadt.de}
\affiliation{ExtreMe Matter Institute EMMI, 
GSI Helmholtzzentrum f\"ur Schwerionenforschung GmbH, 
64291 Darmstadt, Germany}
\affiliation{Institut f\"ur Kernphysik, 
Technische Universit\"at Darmstadt, 
64289 Darmstadt, Germany}

\begin{abstract}
We perform state-of-the-art large-scale shell-model calculations of
the structure factors for elastic spin-dependent WIMP scattering off
$^{129,131}$Xe, $^{127}$I, $^{73}$Ge, $^{19}$F, $^{23}$Na ,$^{27}$Al,
and $^{29}$Si. This comprehensive survey covers the nonzero-spin
nuclei relevant to direct dark matter detection. We include a
pedagogical presentation of the formalism necessary to describe
elastic and inelastic WIMP-nucleus scattering. The valence spaces and
nuclear interactions employed have been previously used in nuclear
structure calculations for these mass regions and yield a good
spectroscopic description of these isotopes. We use spin-dependent
WIMP-nucleus currents based on chiral effective field theory (EFT) at
the one-body level and including the leading long-range two-body
currents due to pion exchange, which are predicted in chiral EFT.
Results for all structure factors are provided with theoretical error
bands due to the nuclear uncertainties of WIMP currents in nuclei.
\end{abstract}


\maketitle

\section{Introduction}

About $25\%$ of the energy density of our Universe consists of dark
matter, a form of matter that rarely interacts with baryons and has
eluded direct observation so far~\cite{Gaitskell,Bertone}. This
large-scale problem is closely connected to new physics at the
smallest scales, because dark matter candidates arise naturally in
extensions of the Standard Model of particle physics~\cite{Feng}.
Prominent dark matter candidates are weakly interacting massive
particles (WIMPs). They are predicted in supersymmetric models, as the
lightest supersymmetric particles (typically neutralinos) and also in
other Standard Model extensions like models with extra dimensions.
WIMPs are especially promising candidates, because they account
naturally for the dark matter energy density established by
observations~\cite{Bertone}. Moreover, WIMPs interact with quarks, and
thus with baryonic matter, opening the door to direct dark matter
detection via elastic scattering off nuclei~\cite{Baudis}. Inferring
properties of dark matter from direct detection therefore requires
detailed knowledge of the structure factors for WIMP scattering off
strongly interacting nuclei.

In this work, we focus on spin-dependent (SD) WIMP
scattering~\cite{IJMPE}, which is relevant because WIMPs can carry
spin. In particular we assume spin $1/2$ WIMPs, such as neutralinos or
other Majorana fermions.  The detection of elastic SD WIMP scattering
has been the goal of several past and ongoing
experiments~\cite{Kims,Picasso,Xenon,Coupp,Simple,Zeplin,XenonSD},
using different nonzero-spin nuclei as target, but so far without
evidence. Evaluating the response of nuclei to WIMPs is
challenging. First, it requires matching the WIMP-quark couplings in a
particular supersymmetric model to WIMP-nucleon currents. Because
quantum chromodynamics (QCD) is nonperturbative at low energies, this
is best achieved using effective
theories~\cite{Haxton1,Cirigliano,MGS,Haxton2}, in which
spin-independent and SD interactions generally enter in leading
order. Second, the WIMP-nucleus response requires reliable
nuclear-structure calculations. This is especially important for SD
interactions, because the response depends on how the spin of the
nucleus is distributed among nucleons (due to attractive interactions
most of the nucleons pair to spin zero). For the isotopes of
interest~\cite{Kims,Picasso,Xenon,Coupp,Simple,Zeplin,XenonSD}, this
involves medium-mass to heavier nuclei and is a challenging many-body
problem. Previous calculations of SD WIMP scattering off
nuclei~\cite{IJMPE,Engel,Ressell_Ge,Ressell_Al,Dean,Divari,Kortelainen,%
Toivanen} have relied on phenomenological WIMP-nucleon currents, and
are based on nuclear-structure calculations that can be improved with
recent advances in nuclear interactions and computing
capabilities. This work presents progress on these fronts.

The typical momentum transfers involved in WIMP scattering off nuclei
are low and of order of the pion mass. In addition, the typical momenta
involved in low-energy nuclear structure are similar. At these momentum
scales, chiral EFT provides a systematic expansion in powers of momenta
$Q$ for nuclear forces and for the coupling to external probes, based on
the symmetries of QCD~\cite{RMP_chiral,RMP_3N}. In addition to the coupling
through one-body (1b) currents, generally at leading order, two-body
(2b) currents enter at higher order and are quantitatively important.

In previous work~\cite{MGS} we have derived the currents for SD WIMP
scattering off nuclei based on chiral EFT, including 1b currents and
the leading long-range 2b currents due to pion exchange, which are
predicted in chiral EFT. As an application, we focused on the
scattering off $^{129,131}$Xe, as they provide the most stringent
limits for WIMP coupling to neutrons~\cite{Xenon100}. Our results have
recently been adopted as benchmark for the XENON100 SD WIMP-nucleon
cross-section limits~\cite{XenonSD}, and have also been used in
Ref.~\cite{Cerdeno}.

More generally, two-body contributions to weak neutral currents have
been shown to be key for providing accurate predictions of
neutrino-deuteron scattering at solar neutrino energies for
SNO~\cite{SNO1,SNO2}. Weak neutral currents based on chiral EFT have
been explored for light nuclei and neutrino breakup in core-collapse
supernovae~\cite{Doron,SN1,SN2,SN3}, and 2b weak charged currents have
been shown to provide important contributions to Gamow-Teller
transitions and double-beta decays of medium-mass nuclei~\cite{Javier}.
Following our previous work~\cite{MGS}, Refs.~\cite{Cannoni,Divari2b}
have reported simple prescriptions to approximately include the
effects of chiral 2b currents in previous calculations of SD
WIMP-nucleus scattering.

This work expands Ref.~\cite{MGS} by presenting state-of-the-art
large-scale shell-model calculations that describe the nonzero-spin
states of all isotopes that are experimentally relevant for SD WIMP
direct detection: $^{129,131}$Xe, $^{127}$I, $^{73}$Ge, $^{19}$F,
$^{23}$Na, $^{27}$Al, and $^{29}$Si. The nuclear-structure
calculations are performed with interactions and valence spaces that
have been tested in these mass regions. Based on the calculated ground
states, we predict the structure factors for elastic SD WIMP
scattering, including chiral 1b and 2b currents with an improved
treatment of the momentum-transfer dependence for higher momentum
transfers. We provide theoretical error bands due to the uncertainties
of WIMP currents in nuclei.

The outline of this article is as follows. In Sec.~\ref{currents}, we
derive the WIMP currents in nuclei based on chiral EFT. All
microscopic inputs needed to compute the structure factors of SD
WIMP-nucleus scattering are discussed in Sec.~\ref{SF}. Combined with
detailed Appendixes, this includes a pedagogical presentation of the
formalism necessary to describe elastic and inelastic WIMP-nucleus
scattering. In Sec.~\ref{results}, we present large-scale
nuclear-structure calculations that describe the nuclei relevant for
SD WIMP direct detection, and compare our results to experiment. We
then calculate the structure factors for elastic SD WIMP scattering
for all cases using chiral EFT currents. We discuss in detail the role
of 2b currents and their uncertainties; the contributions of different
multipole operators to the total response; and the issue of
proton/neutron versus isoscalar/isovector decompositions of the
structure factors. Finally, we summarize in Sec.~\ref{conclusions}
and give an outlook for future improvements of the nuclear physics of
dark matter detection.

\section{WIMP-nucleus interactions}
\label{currents}

\subsection{Chiral EFT and WIMP currents}

At the WIMP-quark level, the low-momentum-transfer Lagrangian density
${\mathcal L}$ for SD interactions is taken to be an
axial-vector--axial-vector coupling~\cite{IJMPE,Kam}:
\begin{align}
{\mathcal L}^{\rm SD}_\chi &= \frac{G_F}{\sqrt{2}} 
\int d^3{\bf r} \, {j}^\mu({\bf r}) {J}^A_\mu({\bf r}) \nonumber \\ 
&= -\frac{G_F}{\sqrt{2}} \int d^3{\bf r} \, \overline\chi {\bm \gamma} \gamma_5
\chi \cdot \sum_q A_q \overline\psi_q {\bm \gamma} \gamma_5 \psi_q \,,
\label{L}
\end{align}
where $G_F$ is the Fermi coupling constant, and ${J}^A_\mu({\bf r})$
and ${j}^\mu({\bf r})$ denote the hadronic current and the leptonic
current of the WIMP, respectively.  $\chi$ is the neutralino field,
$\psi_q$ are the fields of $q=u,d,s$ quarks, and $A_q$ the
neutralino-quark coupling constants.  The temporal components can be
neglected, because the velocities of WIMPs are expected to be
nonrelativistic with $v/c \sim 10^{-3}$. We also neglect contributions
to the Lagrangian density other than axial-vector currents, such as
polar-vector currents, which are suppressed by the momentum transfer
over the nucleon mass $p/m$~\cite{IJMPE}. This approximation will be
studied in a future paper.

For the WIMP-nucleus response, the SD WIMP interaction couples
dominantly to a single nucleon, but also to pairs of nucleons.
At the one-nucleon level, the quark currents are replaced by their
expectation value in the nucleon, leading to 1b axial-vector
currents ${\bf J}_{i,{\rm 1b}}$. In the nucleus, the currents are
summed over all $A$ nucleons:
\begin{equation}
\sum_q A_q \overline\psi_q {\bm \gamma} \gamma_5 \psi_q
\longrightarrow \sum_{i=1}^A {\bf J}_{i,{\rm 1b}}
= \sum_{i=1}^A ( {\bf J}^0_{i,{\rm 1b}} + {\bf J}^3_{i,{\rm 1b}} ) \,,
\end{equation}
with the isoscalar ${\bf J}^0_{i,{\rm 1b}}$ and isovector 
${\bf J}^3_{i,{\rm 1b}}$ parts.

The coupling of the isoscalar part is given by~\cite{IJMPE}
\begin{equation}
a_0= (A_u+A_d)(\Delta u + \Delta d) + 2A_s \Delta s \,,
\label{a0}
\end{equation}
where $\Delta u, \Delta d, \Delta s$ are defined as $\overline\psi_q
{\bm \gamma} \gamma_5 \psi_q=\Delta q \, {\bm \sigma}/2$, with the
nucleon spin ${\bm \sigma}/2$. Therefore, ${\bf J}^0_{i,{\rm 1b}} =
a_0 \, {\bm \sigma}/2$, and $a_0$ receives contributions from the
isoscalar combination of the $u$ and $d$ quarks to the spin of the
nucleon, as well as from the $s$ quark. Analogously, the isovector
coupling can be written as
\begin{equation}
a_1=(A_u-A_d) (\Delta u - \Delta d) = (A_u-A_d) g_A \,,
\end{equation}
where $g_A$ is the axial coupling constant. This shows that the
isovector part ${\bf J}^3_{i,{\rm 1b}}$ of the axial-vector
WIMP-nucleon coupling is identical, up to replacing $a_1$ by $g_A$, to
the axial-vector part of the weak neutral current.

\subsection{Coupling to one nucleon}
\label{singlenucleon}

The weak neutral current was derived within chiral EFT for
calculations of low-energy electroweak reactions. At lowest orders
$Q^0$ and $Q^2$, there are only 1b currents. For the isovector part of
the axial-vector WIMP-nucleon current, this leads to~\cite{MGS}
\begin{equation}
{\bf J}^3_{i,{\rm 1b}} = \frac{1}{2} \,
a_1 \tau_i^3 \biggl( \frac{g_A(p^2)}{g_A} \, {\bm \sigma}_i
- \frac{g_{P}(p^{2})}{2 m g_A} \,
({\bf p} \cdot {\bm \sigma}_{i}) \, {\bf p} \biggl) \,,
\end{equation}
where $\tau_i^3$ denotes the isospin, ${\bf p}={\bf p}_i-{\bf p}_i'$
the momentum transfer from nucleons to neutralinos, and $g_{A}(p^{2})$
and $g_{P}(p^{2})$ the axial and pseudo-scalar couplings. The momentum
transfer dependence of $g_{A}(p^{2})$ and $g_{P}(p^{2})$ is due to
loop corrections and pion propagators. To order $Q^2$, one
has~\cite{Bernard}
\begin{align}
\frac{g_{A}(p^{2})}{g_{A}} &= 1-2 \, \frac{p^2}{\Lambda_{A}^2} \,, \\
g_P(p^2) &= \frac{2 g_{\pi p n} F_\pi}{m_\pi^2+p^2} - 4 \, 
\frac{m g_A}{\Lambda_A^2} \,,
\end{align}
with $\Lambda_A = 1040 \mev$, pion mass $m_{\pi}=138.04 \mev$, pion
decay constant $F_\pi = 92.4 \mev$, and $g_{\pi p n} = 13.05$.
Chiral 1b currents are similar to the currents used in previous
calculations of WIMP scattering off nuclei~\cite{IJMPE}. The
differences are that the $1/\Lambda_A^2$ terms were neglected and
the Goldberger-Treiman relation was implicitly used to write
$\frac{g_P(p^2)}{2 m g_A} \approx \frac{1}{m_\pi^2+p^2}$. Both present
few percent corrections, but the former increases with momentum transfer.

The axial-vector part of the weak neutral currents is isovector in the
Standard Model, neglecting the strange quark contribution to $a_0$ in
Eq.~(\ref{a0}). Therefore, higher-order $Q^2$ contributions to the
isoscalar WIMP-nucleon current ${\bf J}^0_{i,{\rm 1b}} = a_0 \, {\bm
\sigma}/2$ depend on models of currents in the nucleon. To order
$Q^2$, these lead to 1b currents with a form-factor mass-scale $\sim
\Lambda_A$~\cite{isoscalar} and without pion propagator contributions.
Because the isovector $1/\Lambda_A^2$ terms contribute at the few
percent level for the typical momentum transfers in WIMP scattering,
we chose to neglect higher-order isoscalar current contributions, as
opposed to introducing a model dependence at this level.
	
\subsection{Coupling to two nucleons}

At order $Q^3$, 2b currents enter in chiral EFT~\cite{Park}. We
consider their long-range parts due to pion exchange, which are
predicted in chiral EFT, and for medium-mass nuclei were found to
dominate over the short-range parts~\cite{MGS}. Because of their
pion-exchange nature, the axial-vector part of the weak neutral 2b
current is isovector, ${\bf J}_{\rm 2b}=\sum^A_{i<j}{\bf J}^3_{ij}$,
with
\begin{align}
{\bf J}^3_{12} &= -\frac{g_A}{2F^2_\pi} \, (\tau_1\times\tau_2)^3
\Biggl( \frac{{\bm \sigma}_2\cdot{\bf k}_2}{m_\pi^2+k_2^2}
\biggl[ \Bigl(c_4+\frac{1}{4m}\Bigr)({\bm \sigma}_1\times{\bf k}_2)
\nonumber \\
&\quad +i \, \frac{{{\bf p}_1+\bf p}'_1}{4m} 
+\Bigl(\frac{1+\hat{c}_6}{4m}\Bigr)({\bm \sigma}_1\times{\bf q}) 
\biggr] -(1\leftrightarrow2)\Biggr)  \nonumber \\
&\quad -\frac{g_A}{F^2_\pi} \, c_3 \biggl[ \tau_1^3 \, \frac{({\bm \sigma}_1
\cdot{\bf k}_1)\,{\bf k}_1}{m_\pi^2+k_1^2} + \tau_2^3 \, \frac{(
{\bm \sigma}_2\cdot{\bf k}_2)\,{\bf k}_2}{m_\pi^2+k_2^2} \biggl] \,,
\label{2bc}	
\end{align}
where ${\bf k}_i={\bf p}'_i-{\bf p}_i$ and ${\bf q}=-{\bf k}_1- {\bf
k}_2$. This improves the treatment of the momentum-transfer
dependence compared to our previous work~\cite{MGS}, as it does not
make the approximation of low-momentum transfers in the
currents~\cite{Park}. As a result, two momentum transfers appear,
${\bf k}_1$ and ${\bf k}_2$, and also a new term proportional to
$1+\hat{c}_6$, which vanishes in the limit of zero momentum transfer.

As in Ref.~\cite{MGS}, we take into account the normal-ordered
one-body part of chiral 2b currents.  This is obtained by summing the
second nucleon $j$ over occupied states in a spin and isospin
symmetric reference state or core, which we take as a Fermi gas:
${\bf J}^{\rm eff}_{i,{\rm 2b}} =\sum_j(1-P_{ij}) {\bf J}^3_{ij}$. The
exchange operator $P_{ij}$ includes all two-body exchange
contributions. Normal ordering is expected to be a very good
approximation for medium-mass and heavy nuclei, because of phase-space
restrictions of normal Fermi systems at low energies~\cite{Fermi}.

The resulting effective 2b currents ${\bf J}^{\rm eff}_{i,{\rm 2b}}$
are derived in detail in Appendix~\ref{2bderivation}. We find that the
leading long-range 2b currents lead to three different contributions.
First, there is a renormalization of the axial coupling~\cite{MGS},
\begin{widetext}
\vspace{-0.5cm}
\begin{align}
{\bf J}^{\text{eff},\sigma}_{i,\text{2b}}(\rho,{\bf p},{\bf P})
=&-g_A {\bm \sigma}_i \, \frac{\tau^3_i}{2} \, \frac{\rho}{2 F^2_\pi}
\Biggl( \frac{1}{3} \Bigl(-c_3+\frac{1}{4m}\Bigr)
\Bigr[ I_1^{\sigma}(\rho,|{\bf P}-{\bf p}|)+I_1^{\sigma}(\rho,|{\bf P}
+{\bf p}|) \Bigr] \nonumber \\
&+\frac{1}{3} \Bigl(c_4+\frac{1}{4m}\Bigr) \Big[
3I_2^{\sigma}(\rho,|{\bf P}-{\bf p}|)-I_1^{\sigma}(|{\bf P}-{\bf p}|)
+3I_2^{\sigma}(\rho,|{\bf P}+{\bf p}|)-I_1^{\sigma}(|{\bf P}+
{\bf p}|) \Big] \nonumber \\
&+ \Bigl(\frac{1+\hat{c}_6}{12m}\Bigr) \biggl[I_{c_6}(\rho,
|{\bf P}-{\bf p}|) \, \frac{{\bf p}\cdot({\bf P}-{\bf p})}{
({\bf P}-{\bf p})^2}
-I_{c_6}(\rho,|{\bf P}+{\bf p}|) \, \frac{{\bf p}\cdot({\bf P}
+{\bf p})}{({\bf P}+{\bf p})^2} \biggr] \Biggl) \,,
\label{scurrent}
\end{align}
which depends on the density $\rho$, the momentum transfer ${\bf p}$
and the total momentum ${\bf P}={\bf p}_i+{\bf p}_i'$ (due to the the
exchange terms). Such renormalization was also found considering
chiral three-nucleon forces as density-dependent two-body
interactions~\cite{Holt}. Second, there is a contribution to the
pseudo-scalar coupling,
\begin{align}
{\bf J}^{\text{eff},P}_{i,\text{2b}}(\rho,{\bf p},{\bf P}) =& 
-g_A \, \frac{\tau^3_i}{2} \, ({\bf p}\cdot{\bm \sigma}_i) \, {\bf p} 
\, \frac{\rho}{2 F_\pi^2} \Biggl( \frac{4c_3}{m_\pi^2+p^2}
 - \frac{1}{3} \Bigl(c_3+c_4\Bigr) \frac{I^P(\rho,|{\bf P}-{\bf p}|)
+I^P(\rho,|{\bf P}+{\bf p}|)}{p^2} \nonumber \\
&+ \Bigl(\frac{1+\hat{c}_6}{12m}\Bigr) \biggl[
\frac{I_{c_6}(\rho,|{\bf P}-{\bf p}|)}{({\bf P}-{\bf p})^2}
+\frac{I_{c_6}(\rho,|{\bf P}+{\bf p}|)}{({\bf P}+{\bf p})^2}
\biggr] \Biggr) \,,
\label{Pcurrent}
\end{align}
\end{widetext}
and third, chiral 2b currents induce pseudo-scalar-type currents
depending on the total momentum,
\begin{align}
{\bf J}^{\text{eff},P1}_{i,\text{2b}}(\rho,{\bf p},{\bf P}) 
&\sim g_A \, \frac{\tau^3_i}{2} \, ({\bf p}\cdot{\bm \sigma}_i)
\, {\bf P} \,, \label{P1current} \\
{\bf J}^{\text{eff},P2}_{i,\text{2b}}(\rho,{\bf p},{\bf P}) 
&\sim g_A \, \frac{\tau^3_i}{2} \, ({\bf P}\cdot{\bm \sigma}_i)
\, {\bf p} \,, \label{P2current} \\
{\bf J}^{\text{eff},P3}_{i,\text{2b}}(\rho,{\bf p},{\bf P}) 
&\sim g_A \, \frac{\tau^3_i}{2} \, ({\bf P}\cdot{\bm \sigma}_i)
\, {\bf P} \,, \label{P3current}
\end{align}
whose analytical expressions can be found in
Appendix~\ref{2bderivation}. The functions $I^\sigma_1(\rho,Q)$,
$I^\sigma_2(\rho,Q)$, $I^P(\rho,Q)$, and $I_{c_6}(\rho,Q)$ are given by
integrals due to the summation over occupied states in the exchange
terms. They can be evaluated analytically, and the explicit
expressions are given in Appendix~\ref{2bderivation}.

The contributions from 2b currents in Eqs.~\eqref{scurrent}--\eqref{P3current}
depend on the density of the reference state $\rho=2k_{\rm F}^3/(3\pi^2)$
($k_{\rm F}$ is the Fermi momentum) and on the low-energy couplings
$c_3$, $c_4$, and $\hat{c}_6$.
For the density $\rho$ we take the range $\rho=0.10
\dots 0.12 \, {\rm fm}^{-3}$, appropriate for the nuclei considered
(see also Ref.~\cite{Javier}).  The low-energy couplings $c_3$ and
$c_4$ also enter pion-nucleon and nucleon-nucleon interactions and
have been determined from data. Here, we consider the $c_3, c_4$
values from the next-to-next-to-next-to-leading order (N$^3$LO) nucleon-nucleon (NN) potentials of Ref.~\cite{EM} (EM) and
Ref.~\cite{EGM} (EGM), as well as from the NN partial wave analysis
(PWA)~\cite{PWA}. To be conservative, we also consider the estimated
uncertainty in these values expected from higher-order order
contributions, $\delta c_3=-\delta c_4 \approx 1 \gevi$~\cite{RMP_chiral}.
The resulting $c_3$ and $c_4$ values are given in Tables~\ref{P0_test}
and~\ref{delta_a}. We take $\hat{c}_6=5.83$ from Ref.~\cite{c6}.

\begin{table}[b]
\begin{center}
\caption{Comparison of $J^{\sigma}(\rho,P)$, which describes the 
axial contribution at $p=0$ from the normal-ordered one-body part of the
long-range 2b currents, ${\bf J}^{\text{eff},\sigma}_{i,\text{2b}}(\rho,P)
=-g_A{\bm \sigma}_i\frac{\tau^3_i}{2}\frac{\rho}{F^2_\pi}J^{\sigma}(\rho,P)$,
evaluated at the Fermi gas mean value $P^2=6 k_\text{F}^2/5$ and at
$P=0$ for a density $\rho=0.10 \, {\rm fm}^{-3}$. The variation
is shown for all $c_3, c_4$ sets considered, and the
relative variation $\Delta J^\sigma/J^\sigma$ between the Fermi gas 
mean value and $P=0$ is given. The $c_i$ and $J^{\sigma}$ values are
in GeV$^{-1}$.\label{P0_test}}
\begin{tabular*}{0.48\textwidth}{l|c|c|c|c|c}
\hline\hline
& $c_3$ & $c_4$ & $J^\sigma(\rho,P)$ & $J^\sigma(\rho,P=0)$ & $\Delta J^{\sigma}/J^{\sigma}$ \\ \hline
EM & $-3.2$ & $5.4$ & $3.20$ & $2.84$ & $0.11$  \\
EM+$\delta c_i$& $-2.2$ & $4.4$ & $2.57$ & $2.26$ & $0.12$  \\
EGM & $-3.4$ & $3.4$ & $2.29$ & $2.10$ & $0.08$  \\
EGM+$\delta c_i$& $-2.4$ & $2.4$ & $1.66$ & $1.53$ & $0.08$\\
PWA & $-4.78$ & $3.96$ & $2.78$ & $2.59$ & $0.07$  \\
PWA+$\delta c_i$ \, & \, $-3.78$ \, & \, $2.96$ \, & $2.15$ & $2.01$ & $0.06$\\
\hline\hline
\end{tabular*}
\end{center}
\end{table}

In Table~\ref{P0_test}, we study the $P$ dependence of the 2b current
contribution to the axial coupling, which we write as ${\bf J}^{\text{eff},
\sigma}_{i,\text{2b}}(\rho,P) =-g_A{\bm \sigma}_i\frac{\tau^3_i}{2}
\frac{\rho}{F^2_\pi}J^{\sigma}(\rho,P)$, at $p=0$. We compare
$J^{\sigma}(\rho,P)$ for the Fermi gas mean value $P^2=6
k_\text{F}^2/5$ and $P=0$ at a density $\rho=0.10 \, {\rm fm}^{-3}$
and for the different $c_3, c_4$ sets considered. Table~\ref{P0_test}
shows that the $P$ dependence is very weak: $J^{\sigma}(\rho,0)$
varies by less than 12\% over the relevant $P$ range. For other
densities in the range $\rho=0.10...0.12 \, {\rm fm}^{-3}$ this
variation is even smaller. Because 2b currents are a correction to the
leading 1b currents, we therefore set $P=0$ in the axial 2b current
contribution, Eq.~\eqref{scurrent}. As the contributions from
Eqs.~\eqref{P1current}--\eqref{P3current} are expected to be weaker,
we therefore consistently set $P=0$, so that only the standard
pseudo-scalar part, Eq.~\eqref{Pcurrent}, contributes. Finally to
connect to our previous work, for $p=P=0$, both $I^\sigma_1$ and
$I^\sigma_2$ lead to~\cite{MGS}
\begin{align}
I^\sigma(\rho,p=P=0) &\equiv I^\sigma_1(\rho,p=P=0)
 = I^\sigma_2(\rho,p=P=0)\nonumber \\
&=1- \frac{3m_\pi^2}{k_\text{F}^2}
+\frac{3m_\pi^3}{k_\text{F}^3} \arctan\biggl(\frac{k_\text{F}}{m_\pi}\biggr),
\end{align}
where $I^{\sigma}(\rho,p=P=0) = 0.58...0.60$ depends only weakly on
the density in the range $\rho=0.10...0.12 \, {\rm fm}^{-3}$.

\begin{table}[t]
\begin{center}
\caption{Values for all $c_3, c_4$ sets considered of the long-range
2b current contributions $\delta a_i(p=0)$ (axial) and $\delta a_i^P
(p=m_\pi)$ (pseudo-scalar) for the density range $\rho=0.10...0.12 \,
{\rm fm}^{-3}$. The $c_i$ values are in GeV$^{-1}$.\label{delta_a}}
\begin{tabular*}{0.48\textwidth}{l|c|c|c|c}
\hline\hline
& $c_3$ & $c_4$ & $\delta a_1(p=0)$ & \, $\delta a_1^P(p=m_\pi)$ \\ \hline
EM & $-3.2$ & $5.4$ & $-(0.26...0.32)$ & $0.32...0.38$ \\
EM+$\delta c_i$ & $-2.2$ & $4.4$ & $-(0.20...0.25)$ & $0.23...0.27$ \\
EGM & $-3.4$ & $3.4$ & $-(0.19...0.24)$ & $0.33...0.39$ \\
EGM+$\delta c_i$ & $-2.4$ & $2.4$ & $-(0.14...0.17)$ & $0.24...0.28$ \\
PWA & $-4.78$ & $3.96$ & $-(0.23...0.29)$ & $0.45...0.54$ \\
PWA+$\delta c_i$ \, & \, $-3.78$ \, & \, $2.96$ \, & 
\, $-(0.18...0.23)$ \, & $0.36...0.43$ \\
\hline\hline
\end{tabular*}
\end{center}
\end{table}

For $P=0$, the 2b current contribution to the axial part,
Eq.~\eqref{scurrent}, can be written as a momentum- and
density-dependent renormalization $\delta a_1(p)$,
\begin{equation}
{\bf J}^{\text{eff},\sigma}_{i,\text{2b}} = g_A{\bm \sigma}_i
\frac{\tau^3_i}{2} \, \delta a_1(p) \,,\\
\end{equation}
with
\begin{align}
\delta a_1(p) =& -\frac{\rho}{F^2_\pi} \, 
\biggl[ \frac{1}{3} \Bigl(c_4+\frac{1}{4m}\Bigr) 
\Bigl[3I_2^{\sigma}(\rho,p)-I_1^{\sigma}(\rho,p) \Bigr] \nonumber \\[1mm]
&+\frac{1}{3}\Bigl(-c_3+\frac{1}{4m}\Bigr) 
I_1^{\sigma}(\rho,p) - \Bigl( \frac{1+\hat{c}_6}{12m} 
\Bigr) I_{c_6}(\rho,p) \biggl] \,.
\end{align}
Similarly, we write the 2b-current contribution to the pseudo-scalar
coupling, Eq.~(\ref{Pcurrent}), as a momentum- and density-dependent
renormalization $\delta a_1^P(p)$,
\begin{equation}
{\bf J}^{\text{eff},P}_{i,\text{2b}} = g_A \, \frac{\tau^3_i}{2}
\, ({\bf p}\cdot{\bm \sigma}_i) \, {\bf p} \, \frac{\delta a_1^P(p)}{p^2} \,,
\end{equation}
with
\begin{align}
\delta a_1^P(p) &= \frac{\rho}{F^2_\pi} \biggl[ 
\frac{-2c_3 p^2}{m^2_\pi+p^2}+\frac{c_3+c_4}{3} \, I^P(\rho,p) 
\nonumber \\[1mm]
&\quad- \frac{1+\hat{c}_6}{12m} \, I_{c_6}(\rho,p) \biggr] \,.
\end{align}
The ranges of $\delta a_1(p)$ and $\delta a_1^P(p)$ are given in
Table~\ref{delta_a} for the $c_3, c_4$ values and the density range
considered. We find that $\delta a_1(p)$ reduces the axial part of the
current by $14\%...32\%$ at $p=0$.  The momentum transfer
dependence is mild, as the reduction is $16\%...36\%$ at
$p=m_{\pi}$. Moreover, $\delta a_1^P(p)$ increases the pseudo-scalar
part of the current by $23\%...54\%$ at $p=m_{\pi}$.  At lower
momentum transfers this enhancement is weaker, while it is more
significant for higher $p$. These results are consistent with studies
of Gamow-Teller transitions and double-beta decays~\cite{Javier}. As
discussed in Ref.~\cite{MGS}, in addition to the long-range 2b
pion-exchange currents, there are short-range 2b currents for the
isoscalar and isovector parts, which are included as contact terms in
chiral EFT. The isovector short-range 2b parts only lead to small
contributions~\cite{Javier}. Therefore, we neglect short-range 2b
currents at this level, which is also consistent with neglecting
higher-order (short-range) 1b isoscalar currents, see
Sec.~\ref{singlenucleon}.

\subsection{Combined response}
Combining the 1b and the long-range 2b currents to order $Q^3$ in
chiral EFT (replacing $g_A$ by $a_1$ for the latter), the isovector
part of the axial-vector WIMP current at the normal-ordered one-body
level is given by~\cite{MGS}
\begin{align}
{\bf J}^3_{i,{\rm 1b+2b}}
&= \frac{1}{2} \, a_1 \tau_i^3 \Biggl[ \biggl(\frac{g_A(p^2)}{g_A}+
\delta a_1(p) \biggr) \, {\bm \sigma}_i \nonumber \\
&\quad +\biggl(-\frac{g_{P}(p^{2})}{2 m g_A} + \frac{\delta a_1^P(p)}{p^2}\biggr)
\, ({\bf p} \cdot {\bm \sigma}_{i}) \, {\bf p} \, \Biggr] \,.
\label{J3}
\end{align}

\section{WIMP-nucleus scattering and structure factors}
\label{SF}

\subsection{WIMP-nucleus scattering}
The differential cross section for SD WIMP elastic scattering off a
nucleus in the initial state $\ket{i}$ to the final state $\ket{f}$
can be obtained from the low-momentum-transfer Lagrangian density of
Eq.~\eqref{L}. A detailed derivation is performed in
Appendix~\ref{SFderivation}. The final result is~\cite{IJMPE}
\begin{align}
\frac{d\sigma}{dp^2} &= \frac{2}{(2J_i+1) \pi v^2} \sum_ {s_f,s_i} 
\sum_{M_f,M_i} \bigl|\bra{f}{\mathcal L}^{\rm SD}_\chi\ket{i}\bigr|^2
\nonumber \\
&= \frac{8 G^2_F}{(2J_i+1) v^2} \, S_A(p) \,,
\end{align}
where the sum $s_f, s_i = \pm 1/2$ is over neutralino spin
projections, and the sum $M_f, M_i$ is over the projections of the total
angular momentum of the final and initial states $J_f, J_i$,
respectively; $v$ is the WIMP velocity, and $S_A(p)$ the axial-vector
structure factor. The structure factor can be decomposed as a sum over
multipoles $L$ with reduced matrix elements of the longitudinal
${\mathcal L}^5_L$, transverse electric ${\mathcal T}^{\mathrm{el}5}_L$,
and transverse magnetic ${\mathcal T}^{\mathrm{mag}5}_L$ projections of
the axial-vector currents:
\begin{align}
S_A(p) =& \sum_{L \geqslant 0} \bigl|\bra {J_f}\!|{\mathcal L}_L^5|\!
\ket{J_i}\bigr|^2 
 +\sum_{L \geqslant 1} \Bigl(\bigl|\bra{ J_f}\!|
{\mathcal T}_L^{\mathrm{el}5}|\!\ket{J_i}\bigr|^2\nonumber \\
&+\bigl|\bra{ J_f}\!|{\mathcal T}_L^{\mathrm{mag}5}|\!
\ket{ J_i}\bigr|^2\Bigr) \,.
\end{align}
The multipole contributions are obtained from the WIMP-nucleus currents
${\bf J}^A({\bf r})$. At the effective one-body level, chiral 1b and
2b currents lead to (see Appendix~\ref{SFderivation} for the
definition of the multipole operators and details of the derivation)
\begin{align}
{\mathcal L}_{L}^5(p) =& \frac{i}{\sqrt{2L+1}}
\sum^A_{i=1} \frac{1}{2}\biggl[a_0+a_1 \tau^3_i \Bigl(1+\delta a_1(p)\nonumber \\
&-\frac{2g_{\pi pn}F_\pi p^2}{2mg_A
(p^2+m^2_\pi)}+\delta a_1^P(p)\Bigr)\biggr] \nonumber \\[1mm]
&\times \!\Bigl[\sqrt{L+1} M_{L,L+1}(p{\bf r}_i)+
\sqrt{L} M_{L,L-1}(p{\bf r}_i)\Bigr], 
\label{multipoles1}
\end{align}
\begin{align}
{\mathcal T}_{L}^{\mathrm{el}5}(p) =& \frac{i}{\sqrt{2L+1}} \nonumber \\
&\times \sum^A_{i=1}\frac{1}{2}\biggl[a_0+a_1\tau^3_i
\Bigl(1-2\frac{p^2}{\Lambda^2_A}+\delta a_1(p)\Bigr)\biggr] \nonumber \\[1mm]
&\times\Bigl[-\sqrt{L}M_{L,L+1}(p{\bf r}_i)
+\sqrt{L+1} M_{L,L-1}(p{\bf r}_i)\Bigr],
\end{align}
\begin{align}
{\mathcal T}_{L}^{\mathrm{mag}5}(p) 
=&\sum^A_{i=1}\frac{1}{2}\biggl[a_0+a_1\tau^3_i
\Bigl(1-2\frac{p^2}{\Lambda^2_A}+\delta a_1(p)\Bigr)\biggr] \nonumber \\[1mm]
&\times M_{L,L}(p{\bf r}_i) \,.
\label{multipoles3}
\end{align}
The matrix elements of the operator $M_{L,L'}(p {\bf r}_i) = j_{L'}(p r_i) 
[Y_{L'}(\hat{\bf r}_i) \, {\bm \sigma}_i]^L$ (with $L'$ and ${\bm\sigma}$
coupled to $L$) are given in Appendix~\ref{mes}.

\subsection{Parity constraints}
\label{PC}

The different multipoles in
Eqs.~(\ref{multipoles1})--(\ref{multipoles3}) have well-defined
parity $\Pi$, which can be deduced from the definitions given in
Appendix~\ref{SFderivation}, Eqs.~(\ref{multi1})--(\ref{multi3}), and
the transformations under parity of
\begin{align*}
&\Pi({\bm \nabla})=-1 \,, &&\Pi(Y_{LM})=(-1)^L \,, 
&&\Pi({\bf Y}^M_{LL1})=(-1)^L \,,
\end{align*}
and the parity of axial-vector one-body currents $\Pi({\bf J}^A)=+1$.
For elastic scattering, where the initial and final states of the
nucleus are identical ($J=J_i=J_f$), only the multipoles with positive
parity ($\Pi=+1$) contribute to the structure factor, so that we have
\begin{align*}
&\Pi({\mathcal L}_L^5)=(-1)^{L+1}
&&\Rightarrow \quad L \text{ odd} \,, \\
&\Pi({\mathcal T}_L^{\mathrm{el}5})=(-1)^{L+1}
&&\Rightarrow \quad L\text{ odd} \,, \\
&\Pi({\mathcal T}_L^{\mathrm{mag}5})=(-1)^L
&&\Rightarrow \quad L\text{ even} \,.
\end{align*}
Hence, for elastic scattering only the odd-$L$ multipoles of the
longitudinal and transverse electric operators and only the even-$L$
multipoles of the transverse magnetic operator contribute. This is
also the case for inelastic scattering between initial and final
states of the same parity. For inelastic scattering involving
different parity states, the above constraints get reversed.
		
\subsection{Time-reversal constraints}

For elastic scattering, time-reversal invariance also constrains the
multipoles that contribute to the structure factor. We can write the
reduced matrix elements of the sum over one-body operators $O_L(i)$ 
as~\cite{Haxton1}
\begin{align}
\langle J \lVert \, \sum^A_{i=1} O_L(i) \, \rVert J \rangle
\sim \sum_{j,j'}\Psi_{J}(j,j') \Bigl(\langle j\lVert O_L\rVert j'\rangle\nonumber\\
+(-1)^{j-j'}\langle j'\lVert O_L \rVert j\rangle \Bigr) \,,
\label{Treversal}
\end{align}
where $\Psi_{J}(j,j')$ denotes the one-body density matrix, and the
sum is over single-particle total angular momenta $j, j'$ (for
simplicity, we have suppressed the sums over radial quantum numbers
$n,n'$ and orbital angular momenta $l,l'$). Therefore, the symmetry
properties of the matrix elements under exchange of initial and final
states determine the allowed $L$ contributions to elastic
scattering. The relevant operator for SD WIMP-nucleus scattering is
$M_{L,L'}$, whose matrix elements are given in
Appendix~\ref{mes}. They transform as
\begin{align}
&\langle n'l'\frac{1}{2}j'\lVert M_{L,L}(p{\bf r}_i)\rVert nl\frac{1}{2}j\rangle
\nonumber \\
&\quad=(-1)^{j+j'}\langle nl\frac{1}{2}j\lVert M_{L,L}(p{\bf r}_i)
\rVert n'l'\frac{1}{2}j'\rangle \,, \\
&\langle n'l'\frac{1}{2}j'\lVert M_{L,L\pm1}(p{\bf r}_i)\rVert nl\frac{1}{2}j\rangle
\nonumber \\
&\quad=(-1)^{j-j'}\langle nl\frac{1}{2}j\lVert M_{L,L\pm1}(p{\bf r}_i)
\rVert n'l'\frac{1}{2}j'\rangle \,.
\end{align}
Therefore, from Eq.~(\ref{Treversal}) it follows that only the
multipoles with $M_{L,L\pm1}$ contribute to elastic
scattering. Considering the different multipoles in
Eqs.~(\ref{multipoles1})--(\ref{multipoles3}), we thus have
\begin{equation}
\langle J\lVert{\mathcal T}_L^{\mathrm{mag}5}\rVert J\rangle=0 \,,
\end{equation}
so that the transverse magnetic multipoles do not contribute to
elastic scattering.

\subsection{Structure factor for elastic SD scattering}

As a result, the structure factor for elastic SD WIMP scattering off
nuclei is given by~\cite{IJMPE}
\begin{equation}
S_A(p) = \sum_{L \text{ odd}}\Bigl(\bigl| \langle J\lVert{\mathcal L}_L^5(p)\rVert
J\rangle \bigr|^2+\bigl| \langle J\lVert{\mathcal T}_L^{\mathrm{el}5}(p)\rVert J\rangle
\bigr|^2\Bigr) \,,
\end{equation}
and only odd-$L$ longitudinal and electric transverse multipoles contribute.

\section{Results}
\label{results}

\subsection{Spectra}

The calculation of the structure factors requires a reliable
description of the nuclei involved in the scattering process. We
perform state-of-the-art large-scale shell-model calculations of the
nuclear states using the code ANTOINE~\cite{Antoine}. For each
nucleus, we solve the many-body problem in an appropriate valence
space, which depends on the nuclear mass region. In all calculations,
we use nuclear interactions that have been previously employed in
nuclear structure and decay studies. To test the quality
of the structure calculations, we first compare the theoretical with
the experimental spectra for all relevant isotopes.

\begin{figure}[b]
\begin{center}
\includegraphics[width=0.48\textwidth,clip=]{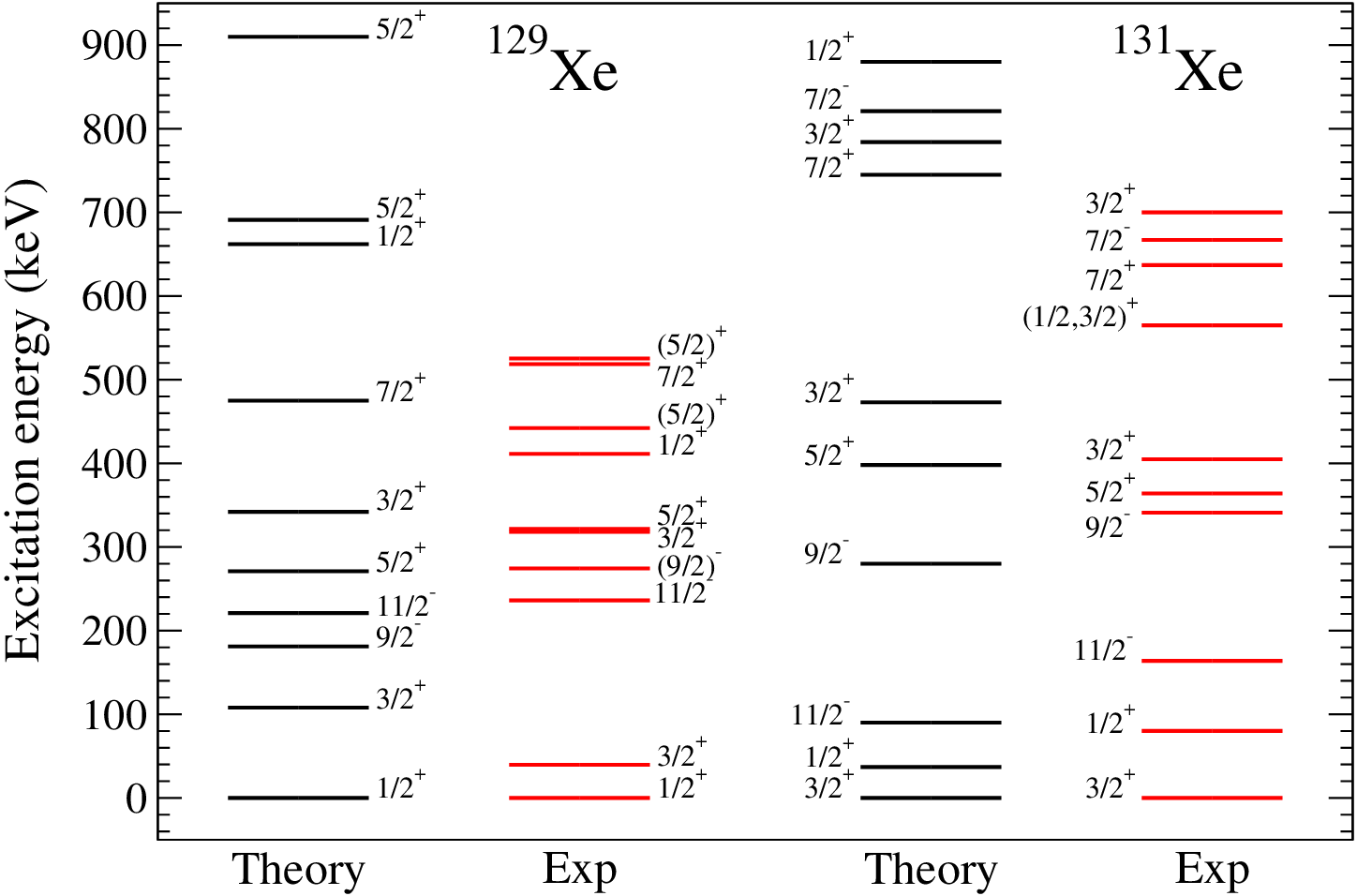}
\end{center}
\caption{(Color online) Comparison of calculated spectra of $^{129}$Xe
and $^{131}$Xe with experiment.\label{spectra}}
\end{figure}

\subsubsection{$^{129}$Xe, $^{131}$Xe, $^{127}$I}

For the heaviest nuclei for SD WIMP scattering, $^{129}$Xe, $^{131}$Xe
and $^{127}$I, the valence space for both protons and neutrons
comprises the $0g_{7/2}$, $1d_{5/2}$, $1d_{3/2}$, $2s_{1/2}$, and
$0h_{11/2}$ orbitals on top of a $^{100}$Sn core.  For $^{131}$Xe we
perform an exact diagonalization in this space.  However, in order to
make the calculations feasible for $^{129}$Xe, the number of particle
excitations from the lower-lying $0g_{7/2}$, $1d_{5/2}$ orbitals into
the $1d_{3/2}$, $2s_{1/2}$, and $0h_{11/2}$ orbitals was limited to
three. With these restrictions the matrix dimension for this space is
$3.5\times 10^8$.  Similarly, for $^{127}$I the number of excitations
into the the $1d_{3/2}$, $2s_{1/2}$, and $0h_{11/2}$ orbitals was
limited to four, leading to a matrix dimension of $4.3\times 10^8$.
For this valence space we have used the so-called GCN5082
interaction~\cite{Menendez_pair, Menendez}, which is based on a
G-matrix with empirical adjustments, mainly in the monopole part, to
describe nuclei within this region.  The same interaction and valence
space have been used to study nuclear structure and double-beta decays
in Refs.~\cite{Menendez_pair,Menendez,Sieja,sm_xenon}.

\begin{figure}[t]
\begin{center}
\includegraphics[width=0.48\textwidth,clip=]{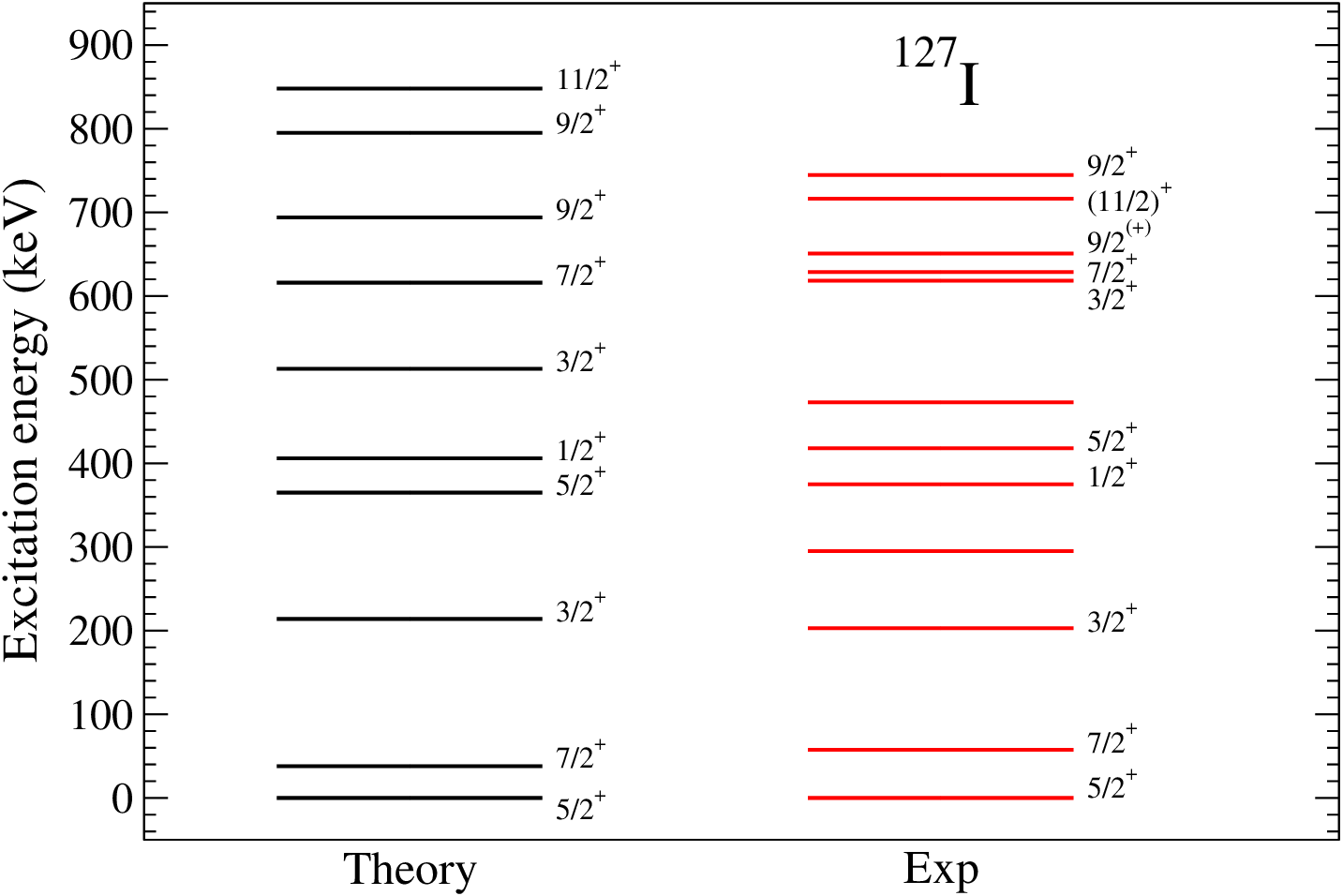}
\end{center}
\caption{(Color online) Comparison of the calculated $^{127}$I spectrum
with experiment.\label{fig:i_spectrum}}
\end{figure}

Figure~\ref{spectra} shows the excitation energies of the lowest-lying
states of $^{129}$Xe and $^{131}$Xe in comparison with experiment (all
energies are measured from the ground state). These spectra have been
previously presented in Ref.~\cite{MGS}. In Fig.~\ref{fig:i_spectrum},
we show the spectrum of $^{127}$I. For all three cases, the
experimental ground state and the overall ordering of the excited
states are very well described. This represents a clear improvement
with respect to previous work~\cite{Toivanen}, and validates the
interaction and valence space used. Note that for $^{127}$I the spin
and parity assignment for some experimental states are not
known. These states are absent in our calculated spectra, which
suggests that they have significant contributions from orbitals lying
outside the valence space considered in the present calculations.

\subsubsection{$^{73}$Ge}

For $^{73}$Ge, the valence space for both protons and neutrons
comprises the $1p_{3/2}$, $0f_{5/2}$, $1p_{1/2}$, and $0g_{9/2}$
orbitals on top of a $^{56}$Ni core. The calculations are performed in
the complete space. We compare results for two different interactions,
the so-called GCN2850 interaction~\cite{Menendez_pair,Menendez}
(Int.~1 in the following) and the RG interaction~\cite{occ76}
(Int.~2). Both are also based on a G-matrix, with mainly monopole
empirical adjustments for this region. They have been employed in beta
and double-beta decay studies, Refs.~\cite{Menendez_pair,Menendez} for
Int.~1 and Refs.~\cite{occ76,zhi} for Int.~2. The former was also used
in a smaller valence space for the description of $^{73}$Ge in
Ref.~\cite{Haxton1}.

\begin{figure}[t]
\begin{center}
\includegraphics[width=0.48\textwidth,clip=]{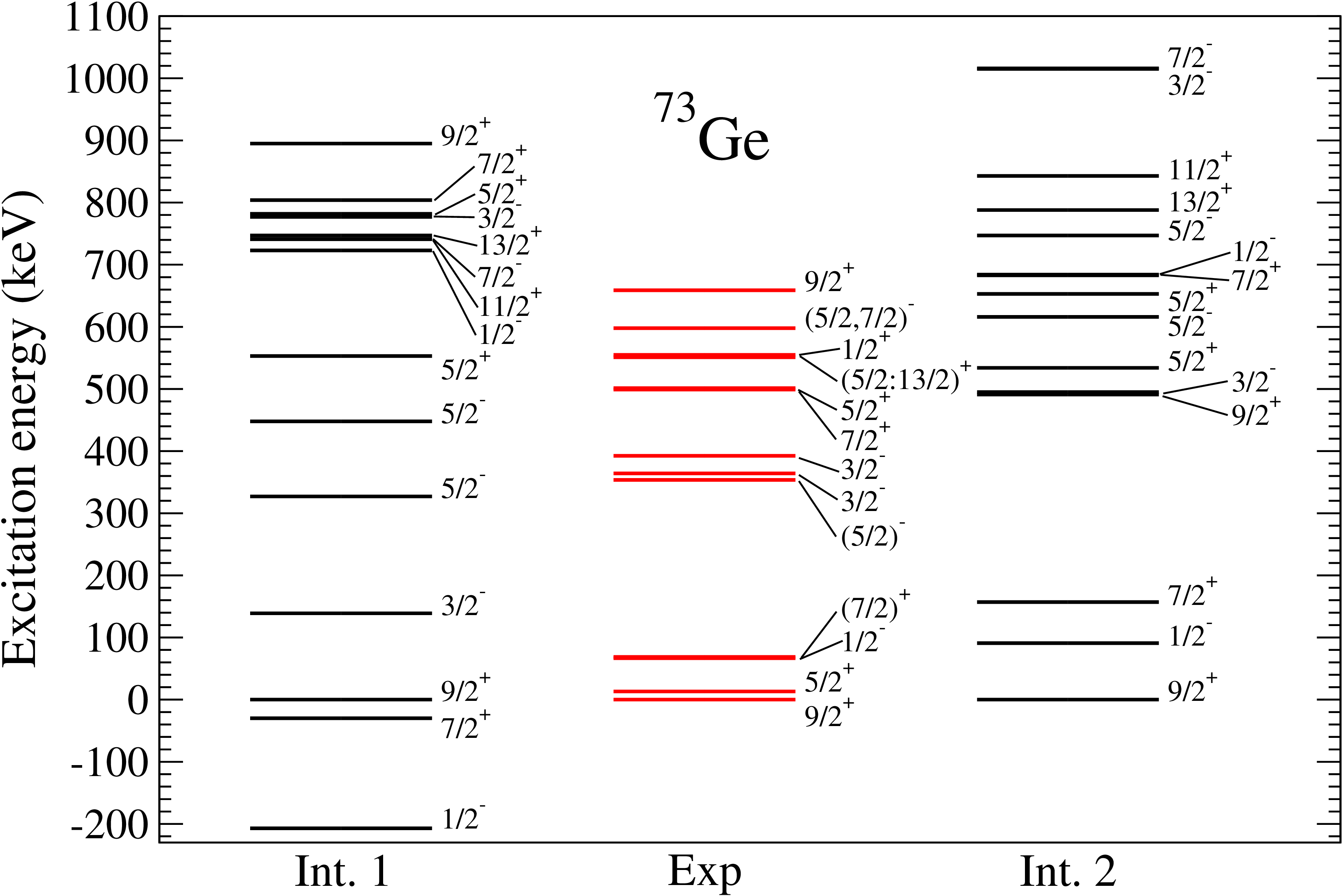}
\end{center}
\caption{(Color online) Comparison of calculated $^{73}$Ge spectra,
using Int.~1 and Int.~2 interactions (for details see text), with 
experiment.\label{fig:ge_spectrum}}
\end{figure}

In Fig.~\ref{fig:ge_spectrum} we compare the resulting spectra with
experiment. We find that the ground state and the overall ordering of
states is much better reproduced by the Int.~2 interaction. In
particular, the structure of three of the lowest-lying states and the
gap between them and the higher-lying states are well described. In
contrast, the Int.~1 interaction predicts a $1/2^-$ ground state, in
disagreement with experiment, and the general spacing of the spectrum
is not well described. Consequently, the Int.~2 interaction will be
the preferred one in this work. Nevertheless, we will also keep the
Int.~1 case, in order to study the sensitivity of the structure factor
to the different nuclear interactions. It is important to note that
the first excited state, which is a $5/2^+$ state, is at too high
excitation energy in both calculations. This suggests that an extended
valence space, probably including the higher-lying $1d_{5/2}$ orbital,
is needed to account for this state. This was also observed in
Ref.~\cite{Ressell_Ge}. A reliable description of the $5/2^+$ state
will be crucial for the study of inelastic scattering off $^{73}$Ge.

\subsubsection{$^{19}$F, $^{23}$Na ,$^{27}$Al, $^{29}$Si}

\begin{figure}[t]
\begin{center}
\includegraphics[width=0.48\textwidth,clip=]{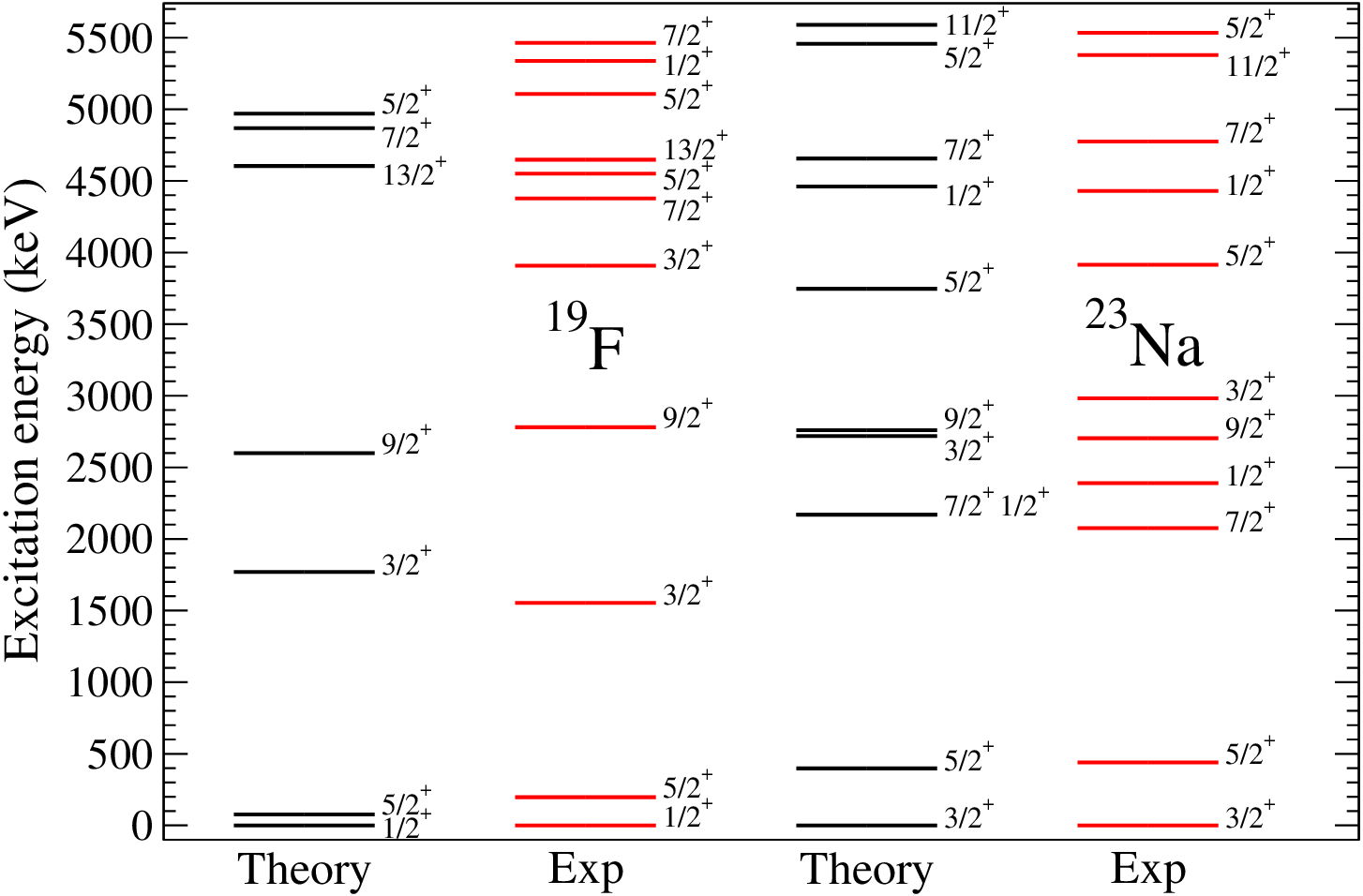}
\end{center}
\caption{(Color online) Comparison of calculated spectra of
$^{19}$F and $^{23}$Na with experiment.\label{fna_spectra}}
\end{figure}

\begin{figure}[t]
\begin{center}
\includegraphics[width=0.48\textwidth,clip=]{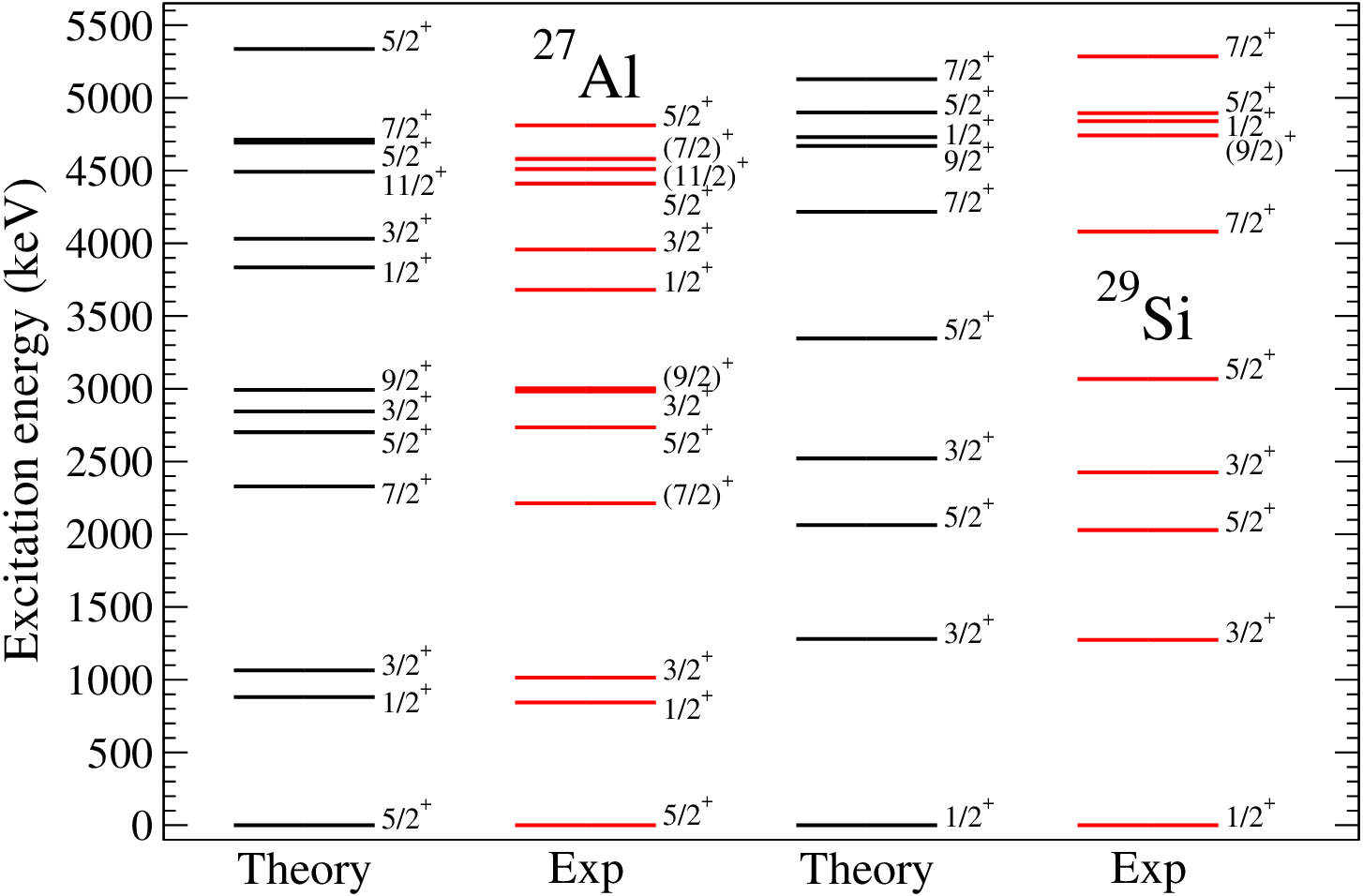}
\end{center}
\caption{(Color online) Comparison of calculated spectra of
$^{27}$Al and $^{29}$Si with experiment.\label{alsi_spectra}}
\end{figure}

\begin{table*}[t]
\begin{center}
\caption{Calculated spin expectation values for protons $\langle {\bf S}_p 
\rangle$ and neutrons $\langle {\bf S}_n \rangle$ of $^{129,131}$Xe, $^{127}$I,
$^{73}$Ge, $^{29}$Si, $^{27}$Al, $^{23}$Na, and $^{19}$F, compared to the
previous calculations of Refs.~\cite{Ressell_Al,Ressell_Ge,Dean,Divari,%
Engel,Toivanen,Kortelainen,Haxton1}.\label{spins}}
\begin{tabular}{l|c|c|c|c|c|c|c|c|c|c|c|c|c|c|c|c}
\hline\hline
& \multicolumn{2}{c|}{$^{129}$Xe} & \multicolumn{2}{c|}{$^{131}$Xe}
& \multicolumn{2}{c|}{$^{127}$I} & \multicolumn{2}{c|}{$^{73}$Ge}
& \multicolumn{2}{c|}{$^{29}$Si} & \multicolumn{2}{c|}{$^{27}$Al}
& \multicolumn{2}{c|}{$^{23}$Na} & \multicolumn{2}{c}{$^{19}$F} \\ \hline
& $\langle {\bf S}_n \rangle$ & $\langle {\bf S}_p \rangle$
& $\langle {\bf S}_n \rangle$ & $\langle {\bf S}_p \rangle$
& $\langle {\bf S}_n \rangle$ & $\langle {\bf S}_p \rangle$
& $\langle {\bf S}_n \rangle$ & $\langle {\bf S}_p \rangle$
& $\langle {\bf S}_n \rangle$ & $\langle {\bf S}_p \rangle$
& $\langle {\bf S}_n \rangle$ & $\langle {\bf S}_p \rangle$
& $\langle {\bf S}_n \rangle$ & $\langle {\bf S}_p \rangle$
& $\langle {\bf S}_n \rangle$ & $\langle {\bf S}_p \rangle$ \\ \hline
This work & $0.329$ & $0.010$ & $-0.272$ & $-0.009$ & $0.031$ 
& $0.342$ & $0.439$ & $0.031$ & $0.156$ & $0.016$ & $0.038$ 
& $0.326$ & $0.024$ & $0.224$ & $-0.002$ & $0.478$ \\
(Int.~1) &&&&&&& $0.450$ & $0.006$ &&&&&&&& \\
\cite{Dean} (Bonn A) & $0.359$ & $0.028$ & $-0.227$ & $-0.009$ 
& $0.075$ & $0.309$ &&&&&&& $0.020$ & $0.248$ && \\
\cite{Dean} (Nijm.~II) & $0.300$ & $0.013$ & $-0.217$ & $-0.012$ 
& $0.064$ & $0.354$ &&&&&&&&&& \\
\cite{Ressell_Al} &&&&&&&&&&& $0.030$ & $0.343$ &&&& \\
\cite{Ressell_Ge} &&&&&&& $0.468$ & $0.011$ & $0.13$ & $-0.002$ &&&&&& \\
\cite{Engel} &&&&&&& $0.378$ & $0.030$ &&&&&&&& \\
\cite{Toivanen}& $0.273$ & $-0.002$ & $-0.125$ & $-7$$\cdot$$10^{-4}$
& $0.030$ & $0.418$ &&&&&&&&&& \\
\cite{Kortelainen} &&&&& $0.038$ & $0.330$ & $0.407$ & $0.005$ 
&&&&& $0.020$ & $0.248$ && \\
\cite{Divari} &  &  &  &  &  & 
 &  &  & $0.133$ & $-0.002$ &&& $0.020$ & $0.248$ & $-0.009$ & $0.475$ \\
\cite{Haxton1} & $0.248$ & $0.007$ & $-0.199$ & $-0.005$ & $0.066$ & 
$0.264$ & $0.475$ & $0.008$ &&&&& $0.020$ & $0.248$ & $-0.009$ & $0.475$ \\
\hline\hline
\end{tabular}
\end{center}
\end{table*}

The valence space of the four lighter nuclei $^{19}$F, $^{23}$Na,
$^{27}$Al, and $^{29}$Si is the $sd$ shell, which comprises the
$0d_{5/2}$, $1s_{1/2}$, and $0d_{3/2}$ orbitals, with a $^{16}$O core.
Full calculations in this valence space are easily performed. In
previous works~\cite{Ressell_Al,Ressell_Ge,Dean,Divari,Haxton1}, the
USD interaction~\cite{usd} was employed. This interaction consists of
a best fit to selected nuclei in this mass region. Here, we use the
more recent USDB interaction~\cite{usdb}, which is an improved version
of USD. The difference between the two interactions is small, see
Sec.~\ref{spinexp}.  In Figs.~\ref{fna_spectra}
and~\ref{alsi_spectra} the positive-parity excited states of all four
nuclei are shown compared to experiment (in the $sd$ shell only
positive-parity states can be obtained). The agreement with experiment
is very good in all cases, both for the ordering and the quantitative
reproduction of the excitation energies.

\subsection{Spin expectation values}
\label{spinexp}

In the limit of low momentum transfer, $p=0$, the structure factor for
elastic SD WIMP scattering is given by the proton and neutron spins
${\bf S}_p=\sum^Z_{i=1}{\bm \sigma}_i/2$ and 
${\bf S}_n=\sum^N_{i=1}{\bm \sigma}_i/2$ in the nucleus~\cite{IJMPE}:
\begin{align}
&S_A(0) = \frac{1}{4 \pi} \bigl|(a_0+a_1') \langle J \lVert {\bf S}_p 
\rVert J\rangle +(a_0-a_1')\langle J\lVert{\bf S}_n\rVert J\rangle
\bigl|^2 \\
&=\frac{(2J+1)(J+1)}{4\pi J} \bigl|(a_0+a_1') \langle {\bf S}_p \rangle
+(a_0-a_1') \langle {\bf S}_n \rangle\bigl|^2 \,,
\label{sf_zero}
\end{align}
where $a_1'=a_1(1+\delta a_1(0))$ includes the effects from chiral 2b
currents. The spin expectation values are defined as $\langle {\bf
S}_{n,p} \rangle = \langle J M=J | {\bf S}^3_{n,p} | J M=J \rangle$.

We list our calculated spin expectation values $\langle {\bf S}_{n,p}
\rangle$ in Table~\ref{spins} in comparison to previous calculations.
As expected for odd-mass nuclei with even number of protons
($^{129,131}$Xe, $^{73}$Ge, and $^{29}$Si) $|\langle {\bf S}_n
\rangle| \gg |\langle {\bf S}_p \rangle|$, while for odd-mass nuclei
with an even number of neutrons ($^{19}$F, $^{23}$Na ,$^{27}$Al, and
$^{127}$I) $|\langle{\bf S}_n \rangle| \ll |\langle {\bf S}_p
\rangle|$. As a result, the WIMP coupling to the even species will be
suppressed. Moreover, the sensitivity to the precise value of the even
species spin is very weak when chiral 2b currents are included. This
is shown in Sec.~\ref{results_SF}. Chiral 2b currents lead to an
interaction of neutrons and protons that overwhelms the direct WIMP
coupling to the suppressed spin expectation value, so that the
structure factors are almost entirely determined by the dominant
$\langle{\bf S}_{n/p} \rangle$ (for odd neutron/proton isotopes).

The spin expectation values of the lighter nuclei, $^{19}$F,
$^{23}$Na, $^{27}$Al, and $^{29}$Si in Table~\ref{spins} are very
close to those of Refs.~\cite{Ressell_Al,Ressell_Ge,Dean,Divari,Haxton1}
due to the similarity of the USD and USDB interactions. This indicates
that the structure for these nuclei is under good control. For
$^{73}$Ge we find a weak sensitivity of the dominant $\langle{\bf S}_n
\rangle$ value comparing the preferred Int.~2 interaction (``This
work'') to the Int.~1 interaction. This range is smaller than the one
in previous calculations of Refs.~\cite{Ressell_Ge,Engel,Kortelainen,Haxton1},
suggesting that the latter may have an even larger variation in the
spectra due to truncations or deficiencies in the interactions
used. Also for the heavier nuclei, $^{129,131}$Xe, and $^{127}$I, we
have performed calculations in the largest spaces to date and with
tested interactions. For $^{129,131}$Xe, the comparison to previous
results is discussed in detail in Ref.~\cite{MGS}. For the dominant
$\langle{\bf S}_n\rangle$ values for $^{129,131}$Xe, and the dominant
$\langle{\bf S}_p\rangle$ value for $^{127}$I, the difference to
previous calculations of Refs.~\cite{Haxton1,Dean,Kortelainen,Toivanen} is
about $25\%$ (and $55\%$ for $^{131}$Xe). We attribute these
differences to the sizable truncations of the valence spaces in those
calculations and because the interactions used have not been as well tested.

\subsection{Structure factors}
\label{results_SF}

\subsubsection{Isoscalar/isovector versus proton/neutron}

The structure factor $S_A(p)$ can be decomposed in terms of its
isoscalar and isovector parts $S_{ij}(p)$, characterized by the
isoscalar and isovector couplings $a_0$ and~$a_1$:
\begin{equation}
S_A(p) = a_0^2 \, S_{00}(p) + a_0 a_1 S_{01}(p) + a_1^2 \, S_{11}(p) \,.
\end{equation}
However, it is common in the literature to use the structure factors
$S_p(p)$ and $S_n(p)$, which are referred to as ``proton-only'' and
``neutron-only'', respectively. They are defined by the couplings
$a_0=a_1=1$ (``proton-only'') and $a_0=-a_1=1$ (``neutron-only'') and
are thus related to the isoscalar and isovector structure factors by
\begin{align}
S_p(p) &= S_{00}(p) +S_{01}(p) +S_{11}(p) \,, \\
S_n(p) &= S_{00}(p) -S_{01}(p) +S_{11}(p) \,.
\end{align}
The origin of the ``proton/neutron-only'' structure factors can be
understood from Eq.~\eqref{sf_zero}. When 2b currents are neglected,
at $p=0$ the ``proton/neutron-only'' structure factors are determined
entirely by the proton/neutron spin expectation values. Moreover, when
the higher-order isovector parts in 1b currents are neglected, this
separation also holds for $p>0$. Because for odd-mass nuclei there is
a clear hierarchy of the spin expectation values (with either
$|\langle {\bf S}_n \rangle| \gg |\langle {\bf S}_p \rangle|$ or
$|\langle {\bf S}_p \rangle| \gg |\langle {\bf S}_n \rangle|$), the
proton/neutron decomposition is useful to capture the dominant parts
of $S_A(p)$. For this reason, and because it is common experimentally,
we will also largely consider the proton/neutron decomposition
here. This is merely a convenient choice of $a_0, a_1$ couplings, but
the notation ``proton/neutron-only'' is misleading, because it does
not imply that the coupling is to protons/neutrons only. Strong
interactions between nucleons in 2b currents, as well as the isovector
nature of pseudo-scalar and other $Q^2$ 1b currents, mean that WIMPs
effectively couple to protons and neutrons in nuclei. In fact, with 2b
currents, both $S_p(p)$ and $S_n(p)$ are determined by the spin
distribution of the odd species.

\begin{figure}[t]
\begin{center}
\includegraphics[width=0.48\textwidth,clip=]{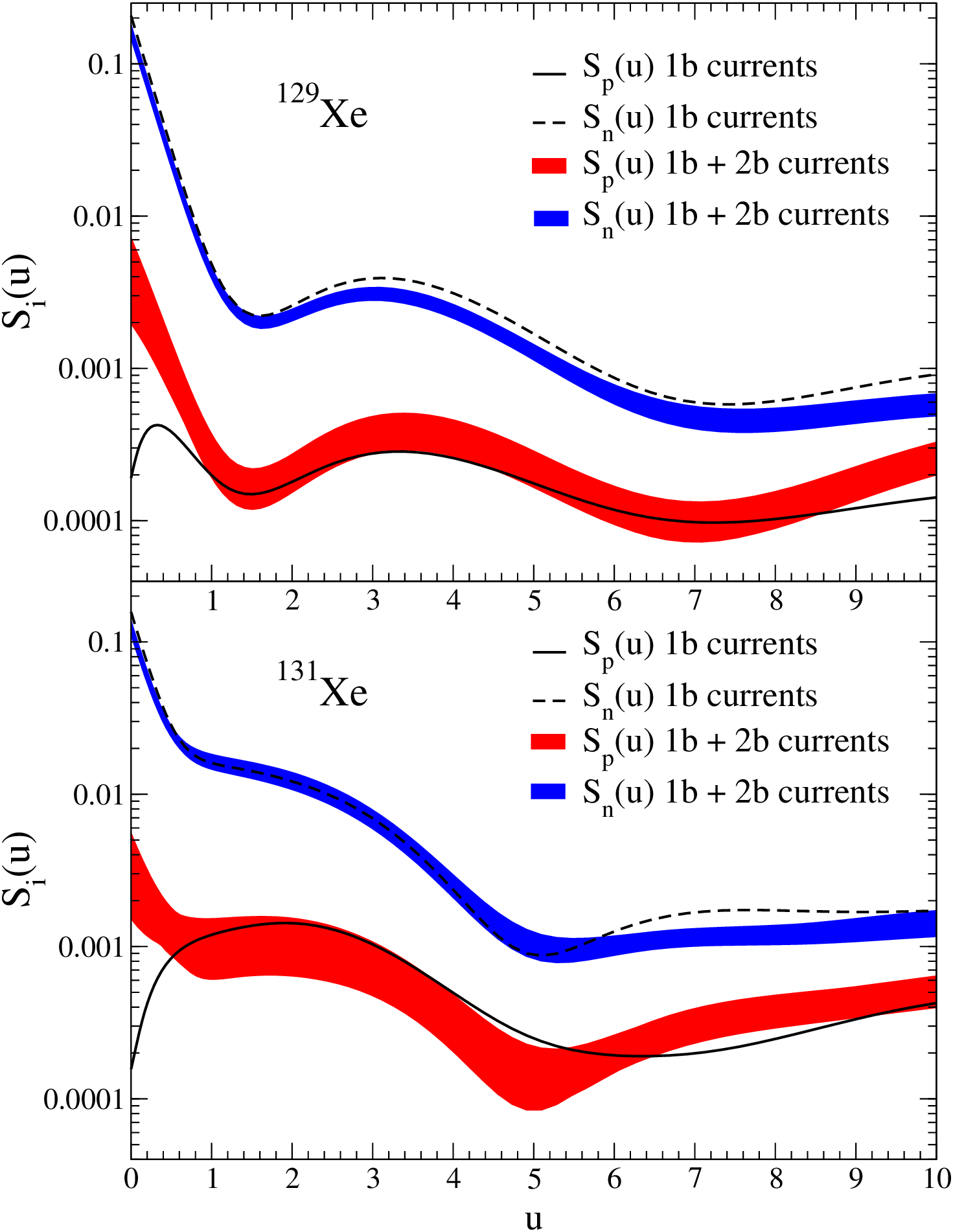}
\end{center}
\caption{(Color online) Structure factors $S_{p}(u)$ (solid lines) 
and $S_{n}(u)$ (dashed) for $^{129}$Xe (top panel) and $^{131}$Xe
(bottom panel) as a function of $u=p^2b^2/2$. The harmonic-oscillator
lengths are $b= 2.2853 \, {\rm fm}$ and $b= 2.2905 \, {\rm fm}$ for
$^{129}$Xe and $^{131}$Xe, respectively. Results are shown at the 
1b current level, and also including 2b currents. The estimated
theoretical uncertainty is given by the red ($S_{p}(u)$) and blue
($S_{n}(u)$) bands.\label{Xe131129_SnSp}}
\end{figure}

\begin{figure}[t]
\begin{center}
\includegraphics[width=0.48\textwidth,clip=]{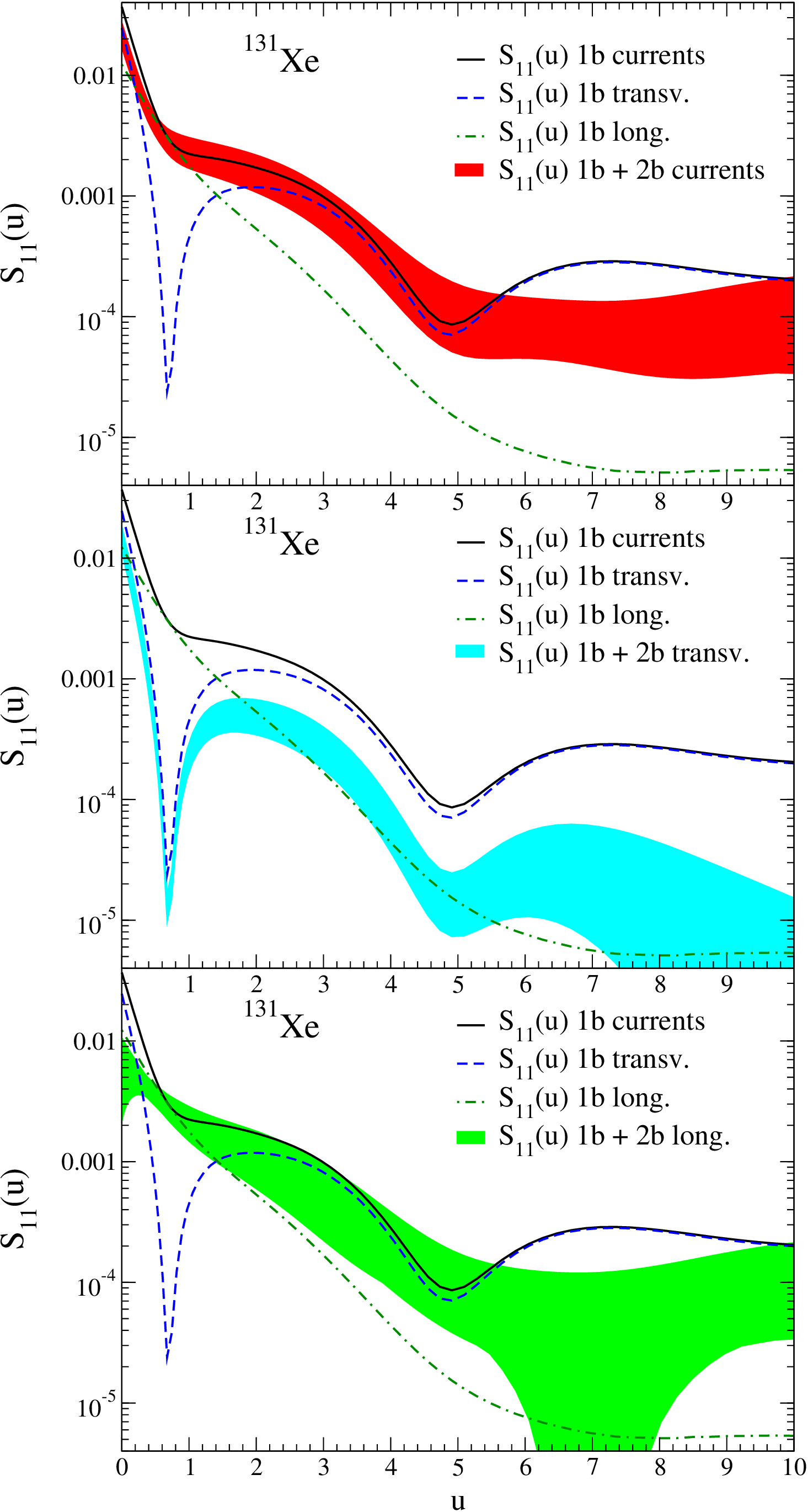}
\end{center}
\caption{(Color online) Decomposition of the isovector structure
factor $S_{11}(u)$ for $^{131}$Xe. At the 1b current level, the full
result (solid, black lines) and the contributions from transverse
electric (dashed, blue) and from longitudinal (dot-dashed, green) multipoles
are shown. The top panel gives also the full 1b plus 2b current
result (red band), while the middle/bottom panels show the
1b plus 2b results when only transverse/longitudinal multipoles
are included (blue/green band). The bands give the estimated 
2b-current uncertainty.\label{Xe131_S11_TL}}
\end{figure}

In the following, we present structure factors as a function of
$u=p^2b^2/2$ with harmonic-oscillator length $b=(\hbar/m\omega)^{1/2}$
and $\hbar \omega=(45 A^{-1/3}-25A^{-2/3}) \, {\rm MeV}$.  When 2b
currents are included, we provide theoretical error bands due to the
uncertainties in WIMP currents in nuclei, see Table~\ref{delta_a}.
This takes into account the uncertainties in the low-energy couplings
$c_3, c_4$ and in the density range $\rho= 0.10...0.12 \fmiq$.

For $^{129}$Xe and $^{131}$Xe the predicted isoscalar/isovector
structure factors $S_{00}(u), S_{01}(u)$, and $S_{11}(u)$ were
discussed in detail in Ref.~\cite{MGS}, and they were compared to the
previous calculations of Refs.~\cite{Dean,Toivanen} (see also
Sec.~\ref{spinexp}). Here, we present in Fig.~\ref{Xe131129_SnSp} the
proton/neutron structure factors $S_p(u)$. At the 1b current level,
the results at $p=0$ are determined by the spin expectation values.
Chiral 2b currents provide important contributions to the structure
factors, especially for $p \lesssim 100 \mev$, where we find in
Fig.~\ref{Xe131129_SnSp} a significant increase of $S_{p}(u)$. This is
because with 2b currents, neutrons can contribute to the
``proton-only'' ($a_0=a_1=1$) coupling due to the axial $\delta a_1(p)$
contribution in Eq.~\eqref{sf_zero}. For $S_n(u)$, 2b currents lead to
a small reduction in the structure factor, depending on the momentum
transfer. This is caused by the combined effect of the axial $\delta
a_1(p)$ and the pseudo-scalar $\delta a_1^P(p)$ contributions. To better
understand how these different contributions enter, we study a
multipole decomposition of the structure factors.

\begin{figure}[t]
\begin{center}
\includegraphics[width=0.48\textwidth,clip=]{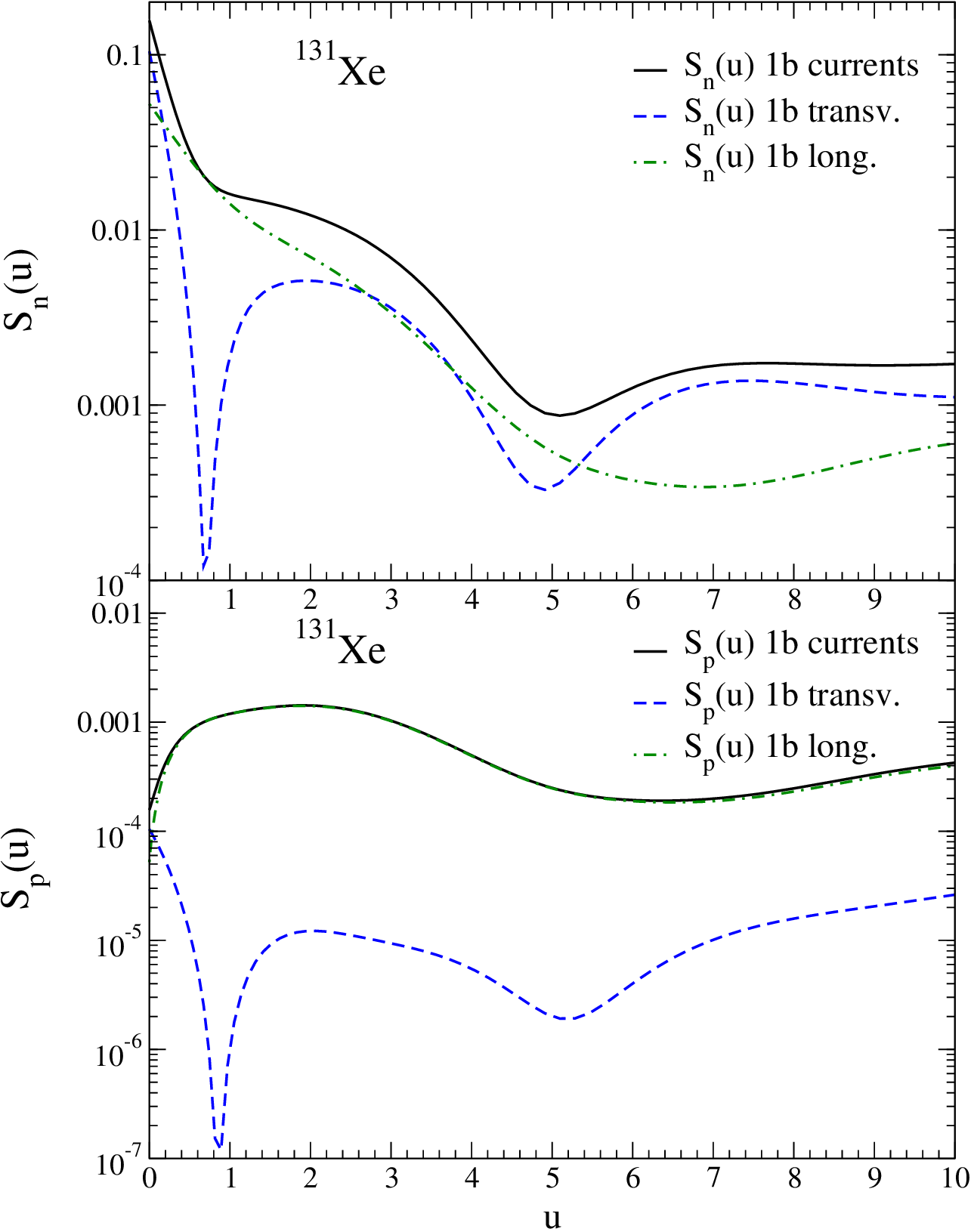}
\end{center}
\caption{(Color online) Structure factors $S_{n}(u)$ (top panel)
and $S_{p}(u)$ (bottom panel) for $^{131}$Xe. At the 1b current level,
the full results (solid, black lines) are compared with the contributions
from transverse electric (dashed, blue) and from longitudinal 
(dot-dashed, green) multipoles.\label{Xe131_Sp_TL}}
\end{figure}

\subsubsection{Multipole decomposition}
\label{multidecomposition}

\begin{figure}[t]
\begin{center}
\includegraphics[width=0.48\textwidth,clip=]{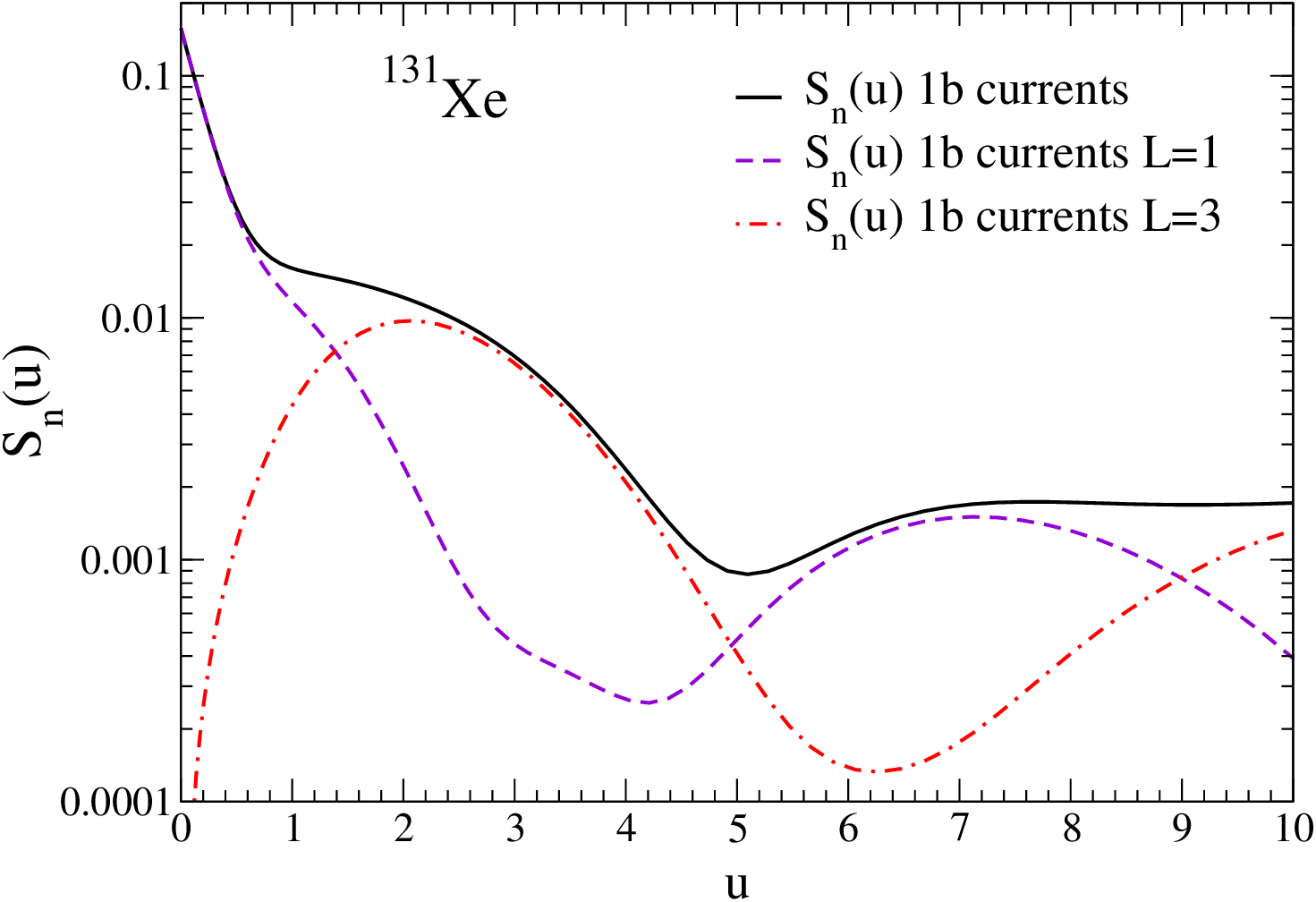}
\end{center}
\caption{(Color online) Decomposition at the 1b current level of the
$^{131}$Xe structure factor $S_{n}(u)$ (solid, black line)
in $L=1$ (dashed, violet) and $L=3$ (dot-dashed, orange) multipoles.
\label{Xe131_Sn_L13}}
\end{figure}

In Fig.~\ref{Xe131_S11_TL} we show the transverse/longitudinal
decomposition of the results with 1b as well as 1b plus 2b currents
for the isovector structure factor $S_{11}(u)$ of $^{131}$Xe (the
long-range 2b currents are isovector). The different 2b current
contributions can be clearly seen in Fig.~\ref{Xe131_S11_TL}. In the
middle panel, where only the transverse electric multipoles are taken
into account, 2b currents reduce the 1b result due to the negative
axial $\delta a_1(p)$ values in Table~\ref{delta_a}. We observe that
the relative reduction depends on $u$ and becomes more important at
higher momentum transfer. The bottom panel shows the longitudinal
multipoles, where both axial $\delta a_1(p)$ and pseudo-scalar $\delta
a_1^P(p)$ 2b current contributions enter. At zero momentum transfer
we find a reduction of the structure factor, driven by $\delta
a_1(p)$, but at $u \sim 0.7$, $p \sim 100 \mev$, this turns into an
enhancement due to $\delta a_1^P(p)$.  In the upper panel, the full 1b
plus 2b band is given, where the final reduction or enhancement over
the 1b result, for a given $u$ value, depends on the relative impact
of the transverse electric and longitudinal multipoles.

It is interesting to study the transverse/longitudinal decomposition
at the 1b level, as shown in Fig.~\ref{Xe131_Sp_TL} for $S_n(u)$ and
$S_p(u)$ of $^{131}$Xe. While both multipoles contribute to $S_n(u)$
(their relative importance depends on $u$), $S_p(u)$ is completely
dominated by the longitudinal multipoles except at $p=0$. In
$^{131}$Xe almost all of the spin is carried by neutrons, so $S_p(0)$
is very small at the 1b level. However, for $p>0$ the (isovector)
pseudo-scalar currents allow neutrons to contribute to $S_p(u)$,
leading to a steep increase in the longitudinal contribution to
$S_p(u)$. Because pseudo-scalar currents only contribute to the
longitudinal multipoles, the transverse part from the protons also
remains very small for $p>0$.

Another way to decompose the structure factors is in terms of the
different $L$ values of the multipoles. Because the ground state of
$^{129}$Xe is $1/2^+$, only $L=1$ contributes. For $^{131}$Xe, with a
$3/2^+$ ground state, $L=1$ and $L=3$ multipoles enter (even-$L$
multipoles are forbidden due to parity, see Sec.~\ref{PC}). The $L$
decomposition of the $^{131}$Xe structure factor $S_{n}(u)$ is shown
in Fig.~\ref{Xe131_Sn_L13}, for simplicity at the 1b current level. We
observe that the $L=3$ multipoles dominate for $1.5 \lesssim u
\lesssim 5$. As a result, the structure factors fall off considerably
more slowly for $^{131}$Xe compared to $^{129}$Xe, where only $L=1$
contributes.

\begin{figure}[t]
\begin{center}
\includegraphics[width=0.48\textwidth,clip=]{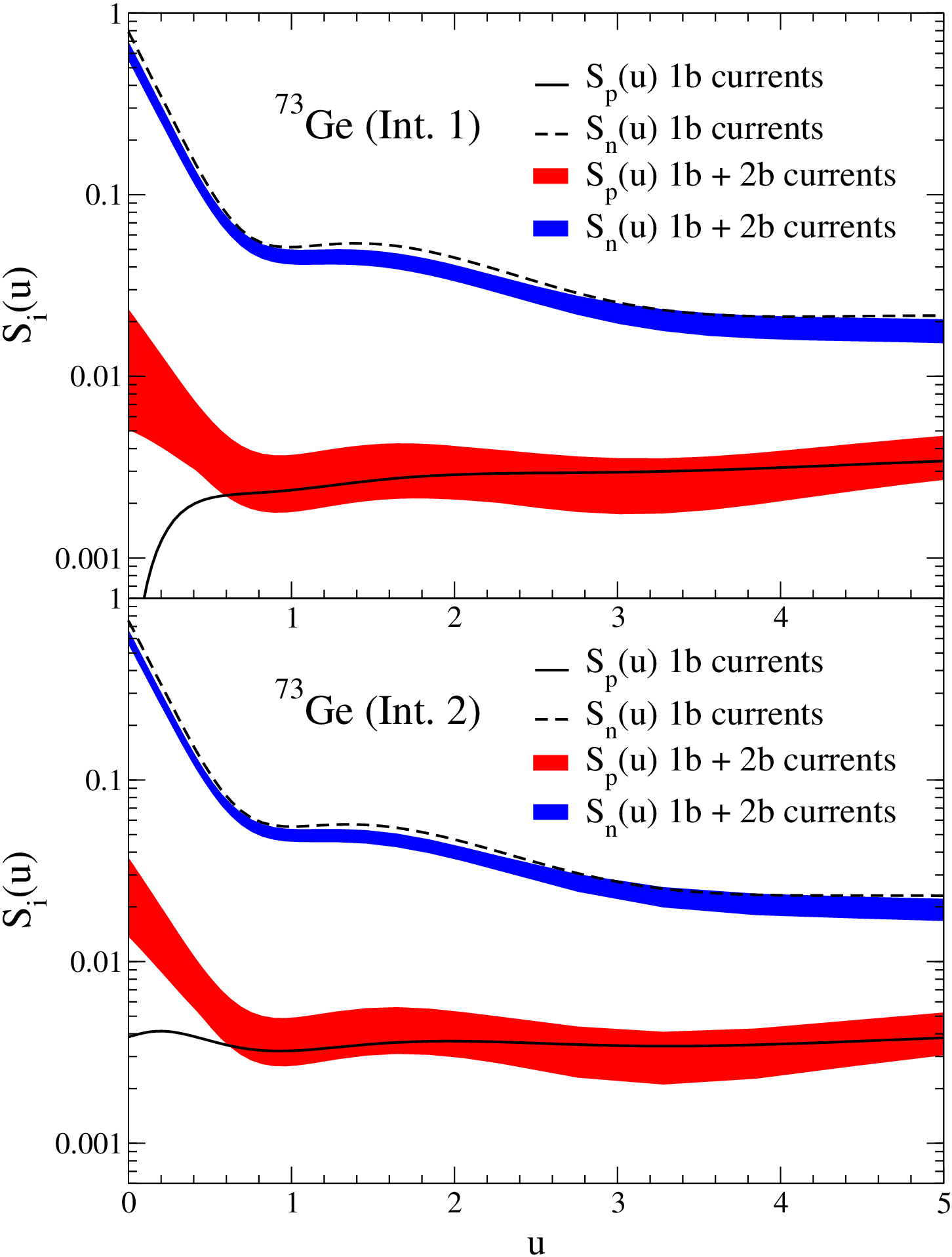}
\end{center}
\caption{(Color online) Structure factors $S_{p}(u)$ (solid lines)
and $S_{n}(u)$ (dashed) for $^{73}$Ge calculated using the Int.~1
(GCN5028, top panel) and the Int.~2 interaction (RG, bottom panel) as
a function of $u=p^2b^2/2$ with $b= 2.1058 \, {\rm fm}$.
Results are shown at the 1b current level, and also including 2b
currents. The estimated theoretical uncertainty is given by the red
($S_{p}(u)$) and blue ($S_{n}(u)$) bands.\label{Ge73_SnSp}}
\end{figure}

\subsubsection{$^{73}$Ge}

Figure~\ref{Ge73_SnSp} shows the structure factors for $^{73}$Ge for
the different Int.~1 and Int.~2 interactions (the latter is preferred
based on the spectra, see Fig.~\ref{fig:ge_spectrum}). The structure
factor $S_n(u)$ differs by less than $10 \%$ between the two
interactions. At the 1b current level, $S_p(u)$ for low momentum
transfers is substantially smaller for Int.~1, due to the very small
$\langle{\bf S}_p\rangle$ value. However, when 2b currents are
included, also for $S_p(u)$ the contributions from neutrons are
dominant, which translates to similar structure factors for the two
interactions. This is because of the similar $\langle{\bf S}_n\rangle$
values (see Table~\ref{spins}) combined with the neutron-proton
coupling through 2b currents.

\begin{figure}[t]
\begin{center}
\includegraphics[width=0.48\textwidth,clip=]{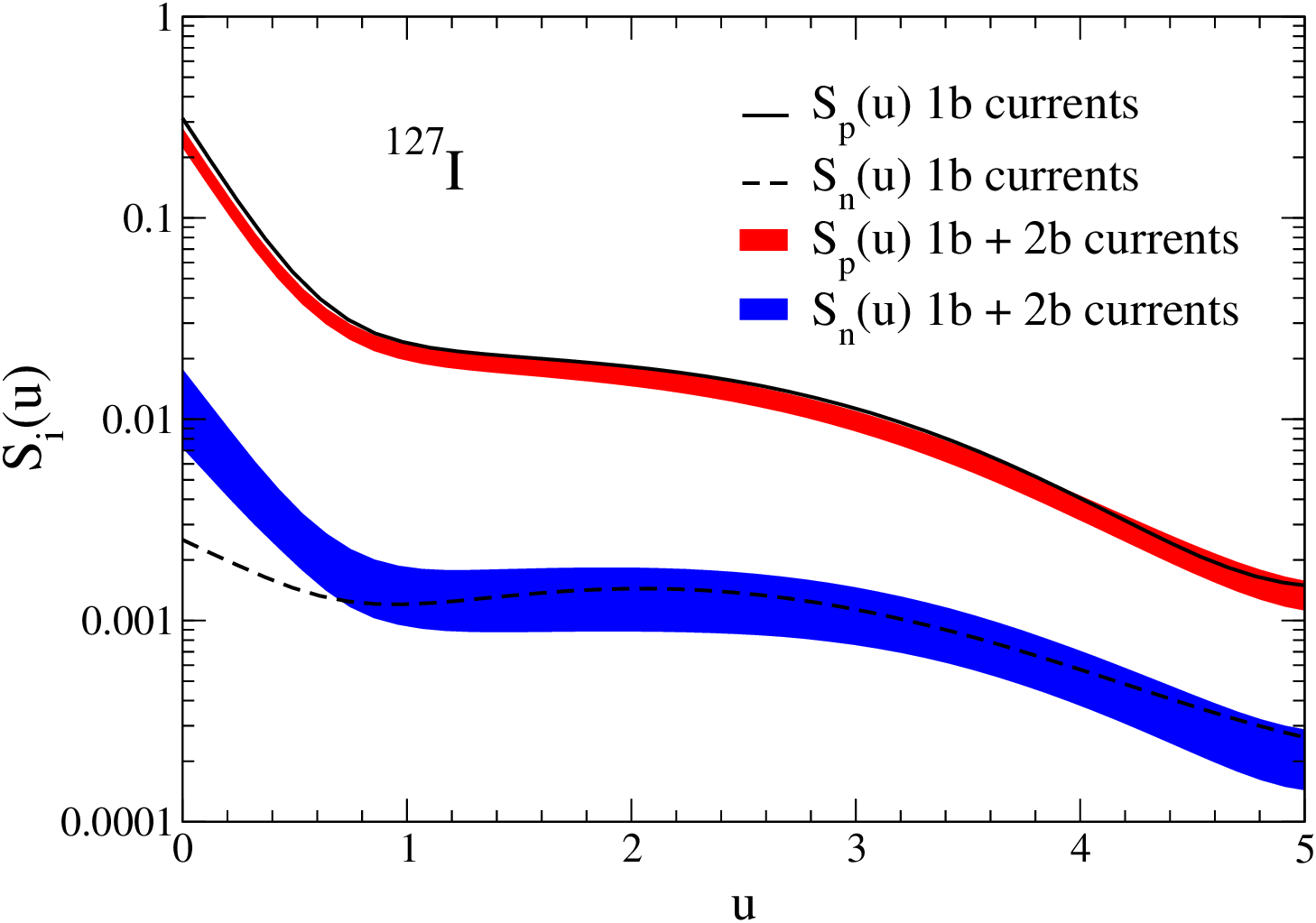}
\end{center}
\caption{(Color online) Structure factors $S_{p}(u)$ (solid lines)
and $S_{n}(u)$ (dashed) for $^{127}$I as a function of $u=p^2b^2/2$
with $b= 2.2801 \, {\rm fm}$. Results are shown at the 1b current
level, and also including 2b currents. The estimated theoretical
uncertainty is given by the red ($S_{p}(u)$) and blue ($S_{n}(u)$)
bands.\label{sf_I}}
\end{figure}

\subsubsection{$^{127}$I, $^{19}$F, $^{23}$Na, $^{27}$Al, $^{29}$Si}

\begin{figure}[t]
\begin{center}
\includegraphics[width=0.48\textwidth,clip=]{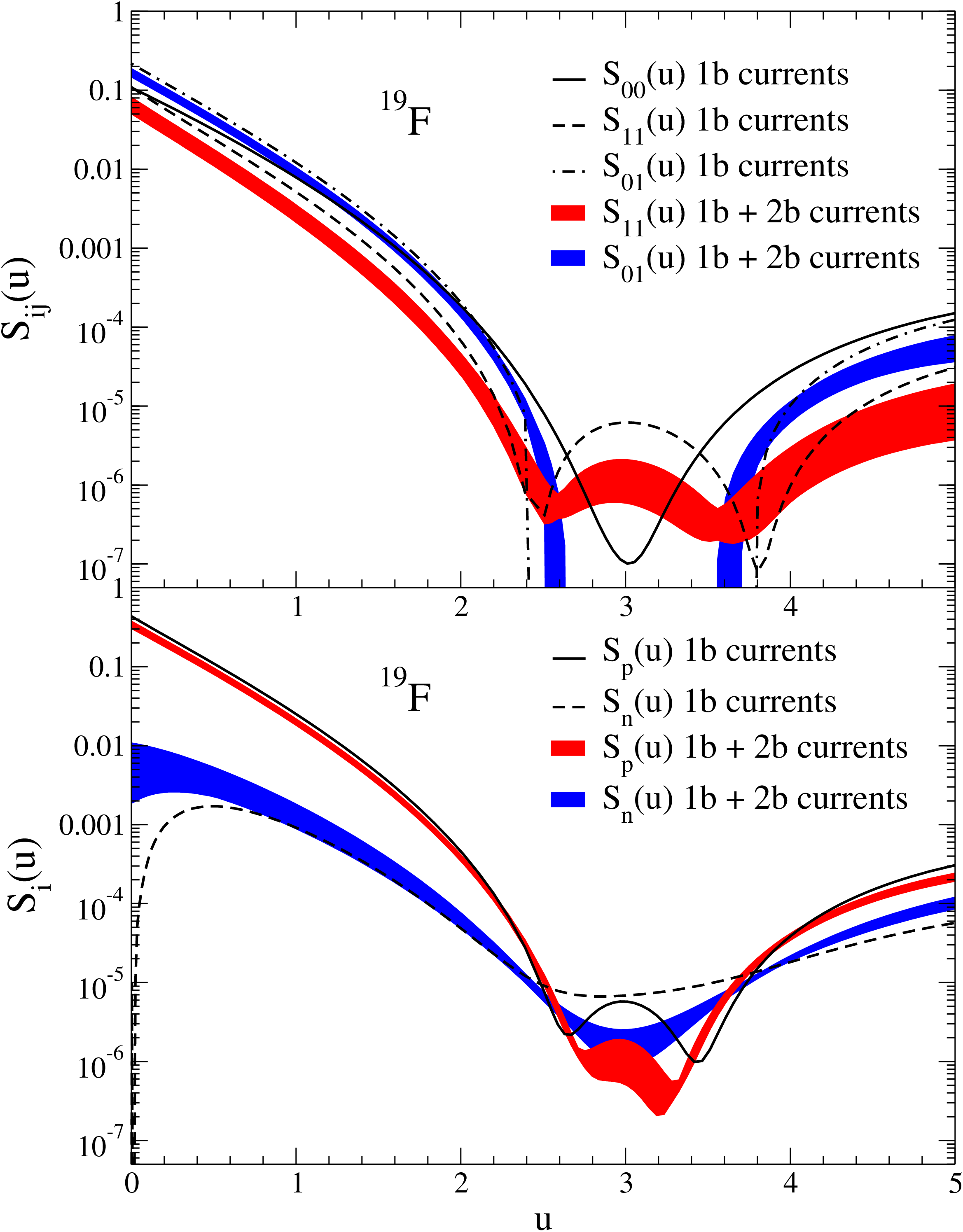}
\end{center}
\caption{(Color online) Structure factors for $^{19}$F as a function
of $u=p^2b^2/2$ with $b= 1.7608 \, {\rm fm}$. Top panel:
Isoscalar/isovector $S_{00}(u)$ (solid line), $S_{01}(u)$ (dashed),
and $S_{11}(u)$ (dot-dashed) decomposition.  Bottom panel:
Proton/neutron $S_{p}(u)$ (solid line) and $S_{n}(u)$ (dashed)
decomposition. In both panels results are shown at the 1b current
level, and also including 2b currents. The estimated theoretical
uncertainty is given by the red ($S_{11}(u)$, $S_{p}(u)$) and blue
($S_{01}(u)$, $S_{n}(u)$) bands.\label{sf_F}}
\end{figure}

\begin{figure}[t]
\begin{center}
\includegraphics[width=0.48\textwidth,clip=]{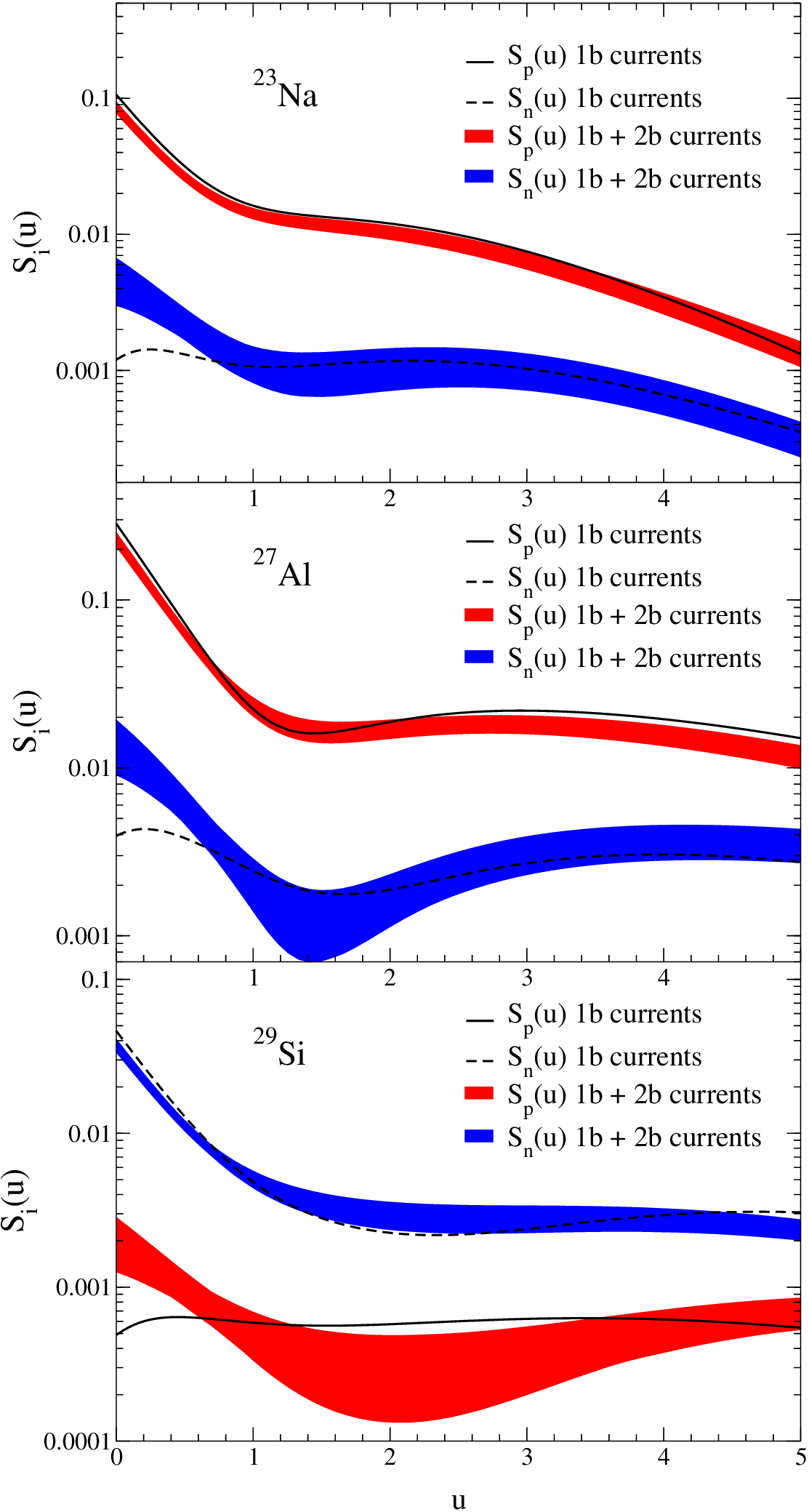}
\end{center}
\caption{(Color online) Structure factors $S_{p}(u)$ (solid lines) 
and $S_{n}(u)$ (dashed) for $^{23}$Na (top panel), $^{27}$Al (middle 
panel), and $^{29}$Si (bottom panel) as a function of $u=p^2b^2/2$, with
harmonic-oscillator lengths $b= 1.8032 \, {\rm fm}$ ($^{23}$Na), 
$b= 1.8405 \, {\rm fm}$ ($^{27}$Al), and $b= 1.8575 \, {\rm fm}$ 
($^{29}$Si). Results are shown at the 1b current level, and also 
including 2b currents. The estimated theoretical uncertainty is 
given by the red ($S_{p}(u)$) and blue ($S_{n}(u)$) bands.\label{sf_NaAlSi}}
\end{figure}

In Figs.~\ref{sf_I},~\ref{sf_F}, and~\ref{sf_NaAlSi}, we show the
structure factors $S_n(u)$ and $S_p(u)$ for $^{127}$I, $^{19}$F,
$^{23}$Na, $^{27}$Al, and $^{29}$Si at the 1b current level and
including 2b currents. The dominant structure factor is the one for
the odd species. Therefore, for $^{29}$Si $S_n(u)$ dominates, while
for the other isotopes $S_p(u)$ is the main component. All the
features discussed for $^{131}$Xe in Sec.~\ref{multidecomposition}
translate to these isotopes as well: The structure factors for the
nondominant ``proton/neutron-only'' couplings are strongly increased
when 2b currents are included. For the dominant structure factor, 2b
currents produce a reduction, by about $10 \%-30 \%$ at low momentum
transfers, which at large $u$ can turn into a weak enhancement due to
the 2b current contribution to the pseudo-scalar currents. This is
most clearly seen for $^{19}$F in the top panel of Fig.~\ref{sf_F},
where we also show the isoscalar/isovector structure factors
$S_{00}(u)$, $S_{01}(u)$, and $S_{11}(u)$. Note that the structure
factor $S_{01}(u)$ vanishes at the point where $S_p(u)$ and $S_n(u)$ cross.

\section{Conclusions and outlook}
\label{conclusions}

This work presents a comprehensive derivation of SD WIMP scattering
off nuclei based on chiral EFT, including one-body currents to order
$Q^2$ and the long-range $Q^3$ two-body currents due to pion exchange,
which are predicted in chiral EFT. Two-body currents are the leading
corrections to the couplings of WIMPs to single nucleons, assumed in
all previous studies. Combined with detailed Appendixes, we have
presented the general formalism necessary to describe both elastic and
inelastic WIMP-nucleus scattering.

We have performed state-of-the-art large-scale shell-model
calculations for all nonzero-spin nuclei relevant to direct dark
matter detection, using the largest valence spaces accessible with
nuclear interactions that have been tested in nuclear structure and
decay studies. The comparison of theoretical and experimental spectra
demonstrate a good description of these isotopes. We have calculated
the structure factors for elastic SD WIMP scattering for all cases
using chiral EFT currents, including theoretical error bands due to
the nuclear uncertainties of WIMP currents in nuclei. Fits for the
structure factors are given in Appendix~\ref{fits}.

We have studied in detail the role of two-body currents, the
contributions of different multipole operators, and the issue of
proton/neutron versus isoscalar/isovector decompositions of the
structure factors. The long-range two-body currents reduce the
isovector parts of the structure factor at low momentum transfer,
while they can lead to a weak enhancement at higher momentum
transfers. Moreover, we have shown that for odd-neutron (odd-proton)
nuclei, two-body currents lead to a significant increase of the
``proton-only'' (``neutron-only'') structure factors, because of
strong interactions between nucleons through two-body currents that
allow the odd species carrying most of the spin to contribute. This
implies that WIMPs effectively couple to protons and neutrons in
nuclei, so that the notation ``proton/neutron-only'' is misleading. In
fact, with 2b currents, both ``proton/neutron-only'' structure factors
are determined by the spin distribution of the odd species.

Future improvements of the nuclear physics of dark matter detection
includes developing shell-model interactions based on chiral EFT,
where the present frontier are semi-magic nuclei up to the calcium
region~\cite{Oxygen,Calcium,Gallant,Wienholtz,proton,Oxygen_spectra},
{\it ab-initio} benchmarks for the lightest isotope $^{19}$F, and expanding
the valences spaces (especially for germanium). In addition, a full
treatment of the one- and two-body currents would require to
renormalize them to the valence space of the many-body calculation,
which can lead to additional contributions to the currents. This and
going beyond the normal-ordering approximation will be pursued in
future work. Moreover, we plan to investigate other
responses~\cite{Haxton1} based on the same large-scale
nuclear-structure calculations presented here.

\section*{Acknowledgements}

We thank L.\ Baudis, J.\ Engel, W.\ C.\ Haxton, R.\ F.\ Lang,
T.\ Marrodan, and T.\ S.\ Park for helpful discussions. This work was
supported by ARCHES, by the DFG through Grant No.~SFB 634, and by the
Helmholtz Alliance Program of the Helmholtz Association, contract
HA216/EMMI ``Extremes of Density and Temperature: Cosmic Matter in the
Laboratory''.

\appendix

\section{Calculation of the\\ effective one-body current 
${\bf J}^{\text{eff}}_{i,\text{2b}}$}
\label{2bderivation}

We calculate the normal-ordered 1b part of 2b currents by summing
the second nucleon over occupied states of a spin and isospin
symmetric reference state or core, which we take as a Fermi gas:
\begin{equation}
{\bf J}^{\text{eff}}_{i,\text{2b}} = \sum_j (1-P_{ij}) {\bf J}^3_{ij} \,,
\end{equation}
where the sum is over occupied states, ${\bf J}^3_{ij}$ is the 2b
current defined in Eq.~\eqref{2bc}, and $P_{ij}$ is the exchange
operator. In this approximation, the momenta ${\bf k}_1$ and 
${\bf k}_2$ in the direct (d) and exchange (ex) contributions are
given by
\begin{align}
{\bf k}_i^{\text{d}} &= {\bf p}'_i-{\bf p}_i = -{\bf p} \,, \\
{\bf k}_i^{\text{ex}} &= P^k_{ij} \, {\bf k}_i^{\text{d}} = {\bf p}_j
-\frac{{\bf P}+{\bf p}}{2} \,, \\
{\bf k}_j^{\text{d}} &= {\bf p}'_j-{\bf p}_j = 0 \,, \\
{\bf k}_j^{\text{ex}} &= P^k_{ij} \, {\bf k}_j^{\text{d}} = -{\bf p}_j
+\frac{{\bf P}-{\bf p}}{2} \,,
\end{align}
where we have used that the initial and final momenta of the nucleon
in the occupied state are identical, ${\bf p}_j={\bf p}'_j$.

The nonvanishing contributions to ${\bf J}^{\text{eff}}_{i,\text{2b}}$
can be grouped into five terms, arising from Eq.~\eqref{2bc}: the
direct (d) and exchange (ex) terms of the $c_3$ term, as well as from the
exchange $c_4$, $p_1$, and $\hat{c}_6$ terms. They read
\begin{widetext}
\begin{align}
&{\bf J}^{\text{eff,d}}_{i,\text{2b}}(c_3\:\text{term})
=-\frac{g_A}{F_\pi^2} \, \frac{\tau^3_i}{2} \,
\frac{8}{(2\pi)^3} \int_0^{k_{\rm F}} \frac{c_3}{m^2_\pi+{\bf p}^2}
\, ({\bf p} \cdot {\bm \sigma}_i) {\bf p} \, d^3{\bf p}_j =-\frac{g_A\rho}{F_\pi^2} \, \frac{\tau^3_i}{2} \, 2c_3 \, \frac{({\bf p}
\cdot{\bm \sigma}_i){\bf p}}{m_\pi^2+{\bf p}^2} \,,
\end{align}
\begin{align}
{\bf J}^\text{eff,ex}_{i,\text{2b}}(c_3\:\text{term})
&=-\frac{g_A}{F_\pi^2} \, \frac{\tau^3_i}{2} \, \frac{4}{(2\pi)^3} 
\frac{1}{2} 
\Biggl[ \int_0^{k_{\rm F}} \frac{c_3}{m_\pi^2+({\bf k}_i^{\text{ex}})^2}
\, ({\bf k}_i^{\text{ex}} \cdot {\bm \sigma}_i) \, 
{\bf k}_i^{\text{ex}} \, d^3{\bf p}_j 
 +\int_0^{k_{\rm F}} \frac{c_3}{m_\pi^2+({\bf k}_j^{\text{ex}})^2}
\, ({\bf k}_j^{\text{ex}} \cdot {\bm \sigma}_i) \, 
{\bf k}_j^{\text{ex}} \, d^3{\bf p}_j \Biggr] \nonumber \\[1mm]
&=-\frac{g_A\rho}{F_\pi^2} \, \frac{\tau^3_i}{2} \, \frac{1}{6} \, c_3
 \biggl[ I_1^{\sigma}(\rho,|{\bf P}-{\bf p}|) \, \bm{\sigma}_i 
+I^P(\rho,|{\bf P}-{\bf p}|) \bigl( (\widehat{{\bf P}-{\bf p}}) \cdot 
\bm{\sigma}_i \bigr) (\widehat{{\bf P}-{\bf p}}) 
+I_1^{\sigma}(\rho,|{\bf P}+{\bf p}|) \, \bm{\sigma}_i \nonumber \\
&\qquad+I^P(\rho,|{\bf P}+{\bf p}|) \bigl( (\widehat{{\bf P}+{\bf p}})\cdot
\bm{\sigma}_i \bigr) (\widehat{{\bf P}+{\bf p}}) \biggr] \,,
\end{align}
\begin{align}
{\bf J}^{\text{eff,ex}}_{i,\text{2b}}(c_4\:\text{term})
&=\frac{g_A}{F_\pi^2} \, \frac{\tau^3_i}{2} \,
\Bigl(c_4+\frac{1}{4m}\Bigr) \frac{4}{(2\pi)^3} \frac{1}{2}
\Biggl[ \int_0^{k_{\rm F}} \frac{1}{m^2_\pi+({\bf k}_i^{\text{ex}})^2}
\, {\bf k}_i^{\text{ex}} \times ({\bm \sigma}_i \times {\bf k}_i^{\text{ex}})
\, d^3{\bf p}_j \nonumber \\
&\qquad +\int_0^{k_{\rm F}} \frac{1}{m^2_\pi+({\bf k}_j^{\text{ex}})^2}
\, {\bf k}_j^{\text{ex}} \times ({\bm \sigma}_i \times {\bf k}_j^{\text{ex}})
\, d^3{\bf p}_j \Biggr] \nonumber \\[1mm]
&=\frac{g_A\rho }{F_\pi^2} \, \frac{\tau^3_i}{2} \, \frac{1}{6}
\Bigl(c_4+\frac{1}{4m}\Bigr) 
\bigg[ \Bigl( 3I_2^{\sigma}(\rho,|{\bf P}-{\bf p}|)
-I_1^{\sigma}(\rho,|{\bf P}-{\bf p}|) 
+3I_2^{\sigma}(\rho,|{\bf P}+{\bf p}|)
-I_1^{\sigma}(\rho,|{\bf P}+{\bf p}|) \Bigr) \, \bm{\sigma}_i \nonumber \\ 
&\qquad -I^P(\rho,|{\bf P}-{\bf p}|) \bigl( (\widehat{{\bf P}-{\bf p}})\cdot
\bm{\sigma}_i \bigr) (\widehat{{\bf P}-{\bf p}})
-I^P(\rho,|{\bf P}+{\bf p}|) \bigl( (\widehat{{\bf P}+{\bf p}})\cdot
\bm{\sigma}_i \bigr) (\widehat{{\bf P}+{\bf p}}) \biggr] \,,
\end{align}
\begin{align}
{\bf J}^\text{eff,ex}_{i,\text{2b}}(p_1\:\text{term})
=&\frac{g_A}{mF_\pi^2} \, \frac{\tau^3_i}{2} \, \frac{4}{(2\pi)^3} 
\frac{1}{8} 
\Biggl[ \int_0^{k_{\rm F}} \frac{\left({\bf p}_j
+\frac{({\bf P}-{\bf p})}{2}\right) \left(\bm{\sigma}_i\cdot{\bf k}_i^{ex}
\right)}{m_{\pi}^2+({\bf k}_i^{ex})^2} \, d^3{\bf p}_j
-\int_0^{k_{\rm F}} \frac{\left({\bf p}_j+\frac{({\bf P}+{\bf p})}{2}
\right)\left(\bm{\sigma}_i\cdot{\bf k}_j^{ex}\right)}{m_{\pi}^2+
({\bf k}_j^{ex})^2} \, d^3{\bf p}_j \Biggr] \nonumber \\[1mm]
=&\frac{g_A \rho}{m F^2_\pi} \, \frac{\tau_i}{2} \, \frac{1}{24} 
\biggl[ I_1^{\sigma}(\rho,|{\bf P}-{\bf p}|) \, \bm{\sigma}_i 
+ I^P_1(\rho,|{\bf P}-{\bf p}|) \bigl( (\widehat{{\bf P}-{\bf p}})\cdot
\bm{\sigma}_i \bigr) (\widehat{{\bf P}-{\bf p}}) \nonumber \\[1mm]
&+\frac{|{\bf p}|}{|{\bf P}-{\bf p}|}I^P_2(\rho,|{\bf P}-{\bf p}|) \bigl( (\widehat{{\bf P}-{\bf p}})\cdot
\bm{\sigma}_i \bigr) (\widehat{{\bf p}}) 
-\frac{1}{4}\frac{|{\bf P}+{\bf p}|}{|{\bf P}-{\bf p}|} I^P_4(\rho,|{\bf P}-{\bf p}|) \bigl( 
(\widehat{{\bf P}-{\bf p}})\cdot\bm{\sigma}_i \bigr) (\widehat{{\bf P}+{\bf p}}) \nonumber \\[1mm]
&+I_1^{\sigma}(\rho,|{\bf P}+{\bf p}|) \, \bm{\sigma}_i 
+I^P_1(\rho,|{\bf P}+{\bf p}|) \bigl( (\widehat{{\bf P}+{\bf p}})
\cdot\bm{\sigma}_i \bigr) (\widehat{{\bf P}+{\bf p}}) \nonumber \\[1mm]
& -\frac{|{\bf p}|}{|{\bf P}+{\bf p}|}I^P_2(\rho,|{\bf P}+{\bf p}|) \bigl( (\widehat{{\bf P}+{\bf p}})\cdot
\bm{\sigma}_i \bigr) (\widehat{{\bf p}})
-\frac{1}{4} \frac{|{\bf P}-{\bf p}|}{|{\bf P}+{\bf p}|} I^P_4(\rho,|{\bf P}+{\bf p}|) \bigl( 
(\widehat{{\bf P}+{\bf p}})\cdot\bm{\sigma}_i \bigr) 
(\widehat{{\bf P}-{\bf p}}) \biggr] \,,
\end{align}
\begin{align}
{\bf J}^\text{eff,ex}_{i,\text{2b}}(\hat{c}_6\:\text{term}) 
&=-\frac{g_A}{F_\pi^2} \, \frac{\tau^3_i}{2} \,
\Bigl(\frac{1+ \hat{c}_6}{4m}\Bigr) \frac{4}{(2\pi)^3} \frac{1}{2} 
\Biggl[ \int_0^{k_{\rm F}} \frac{{\bf p}\times
\left(\bm{\sigma}_i\times{\bf k}_i^{ex}+i{\bf k}_i^{ex}\right)}{m_{\pi}^2
+({\bf k}_i^{ex})^2} \, d^3{\bf p}_j
 + \int_0^{k_{\rm F}} \frac{{\bf p}\times\left(\bm{\sigma}_i\times
{\bf k}_j^{ex}-i{\bf k}_j^{ex}\right)}{m_{\pi}^2+({\bf k}_j^{ex})^2} \, 
d^3{\bf p}_j \Biggr] \nonumber \\[1mm]
&=\frac{g_A \rho}{m F^2_\pi} \, \frac{\tau_2}{2} \, 
\frac{1+\hat{c}_6}{4} \frac{1}{6} 
\Biggl[ \frac{I_{c_6}(\rho,|{\bf P}-{\bf p}|) \, {\bf p}\times
\left(\bm{\sigma}_i\times({\bf P}-{\bf p})\right)}{({\bf P}-{\bf p})^2} 
- \frac{i I_{c_6}(\rho,|{\bf P}-{\bf p}|) \, {\bf p}\times{\bf P}}{
({\bf P}-{\bf p})^2} \nonumber \\[1mm]
&\qquad - \frac{I_{c_6}(\rho,|{\bf P}+{\bf p}|) \, {\bf p}\times\left(
\bm{\sigma}_i\times({\bf P}+{\bf p})\right)}{({\bf P}+{\bf p})^2}
- \frac{i I_{c_6}(\rho,|{\bf P}+{\bf p}|) \, {\bf p}\times{\bf P}}{
({\bf P}+{\bf p})^2} \Biggr] \,,
\end{align}
where the integrals
$I^\sigma_1(\rho,Q)$, $I^\sigma_2(\rho,Q)$, $I^P(\rho,Q)$,
$I^P_{1,2,4}(\rho,Q)$, and $I_{c_6}(\rho,Q)$ are given by the
following expressions
\begin{align}
I^\sigma_1(\rho,Q)
&=\frac{1}{k_\text{F}^3}\frac{9}{4}
\int^{k_\text{F}}_0 \int^1_{-1} \frac{p^4(1-\cos^2\theta)}{
m_\pi^2+p^2+\frac{Q^2}{4}-pQ\cos\theta} \, dp \, d\cos\theta
\nonumber \\[1mm]
&=\frac{1}{512 k_\text{F}^3 Q^3} \Biggl(
8k_\text{F} Q \Bigl[48 (k_\text{F}^2+m_\pi^2)^2 
+32 (k_\text{F}^2-3m_\pi^2) Q^2-3Q^4\Bigr] 
+768 m_\pi^3Q^3\text{arccot}\biggl[
\frac{m_\pi^2+\frac{Q^2}{4}-k_\text{F}^2}{2m_\pi k_\text{F}}\biggr]\nonumber \\
&\quad+3 \Bigl[16 (k_\text{F}^2+m_\pi^2)^2-8 (k_\text{F}^2-5m_\pi^2)
Q^2+Q^4 \Bigr]
\Bigl[4 (k_\text{F}^2+m_\pi^2)-Q^2\Bigr] 
\log\biggl[\frac{m_\pi^2+\bigl(k_\text{F}-\frac{Q}{2}\bigr)^2}{
m_\pi^2+\bigl(k_\text{F}+\frac{Q}{2}\bigr)^2}\biggr] \Biggr) \,,
\end{align}
\begin{align}
&I^\sigma_2(\rho,Q)
=\frac{1}{k_\text{F}^3}\frac{3}{2} \int^{k_\text{F}}_0\int^1_{-1}
\frac{p^4+\frac{p^2 Q^2}{4}-p^3 Q \cos\theta}{
m_\pi^2+p^2+\frac{Q^2}{4}-pQ\cos\theta} \, dp \, d\cos\theta
\nonumber \\[1mm]
&=\frac{1}{16k_\text{F}^3Q}\Biggl( 8k_\text{F} (2k_\text{F}^2-3m_\pi^2) Q
+24 m_\pi^3Q \, \text{arccot}\biggl[\frac{
m_\pi^2+\frac{Q^2}{4}-k_\text{F}^2}{2m_\pi k_\text{F}}\biggr] 
+3m_\pi^2 \Bigl[4k_\text{F}^2-Q^2+4m_\pi^2\Bigr]
\log\biggl[\frac{m_\pi^2+\bigl(k_\text{F}-\frac{Q}{2}\bigr)^2}{
m_\pi^2+\bigl(k_\text{F}+\frac{Q}{2}\bigr)^2}\biggr] \Biggr) \,,
\end{align}
\begin{align}
I^P(\rho,Q)
&=\frac{1}{k_\text{F}^3}\frac{9}{8} \int^{k_\text{F}}_0 \int^1_{-1}
\frac{p^4(6\cos^2\theta-2)-4p^3Q\cos\theta+p^2Q^2}{
m_\pi^2+p^2+\frac{Q^2}{4}-pQ\cos\theta} \, dp \, d\cos\theta \nonumber \\[1mm]
&= -\frac{3}{512 k_{\rm F}^3 Q^3} \Biggl( 8k_{\rm F}Q \Bigl[ 
48(k_{\rm F}^2+m_{\pi}^2)^2 %\nonumber \\ %&\quad 
-32k_{\rm F}^2Q^2-3Q^4 \Bigr]
 + 3 \Bigl[ 4(k_{\rm F}^2+m_{\pi}^2)-Q^2 \Bigr]
\Bigl[4m_{\pi}^2+(2k_{\rm F}-Q)^2 \Bigr] \nonumber \\
&\quad \times \Bigl[ 4m_{\pi}^2+(2k_{\rm F}+Q)^2\Bigr]
%\log\biggl[\frac{4m_\pi^2+(2k_{\rm F}-Q)^2)}{4m_\pi^2+(2k_{\rm F}+Q)^2)}
%\biggr] \Biggr) \,,
\log\biggl[\frac{m_\pi^2+\bigl(k_\text{F}-\frac{Q}{2}\bigr)^2}{
m_\pi^2+\bigl(k_\text{F}+\frac{Q}{2}\bigr)^2}\biggr] \Biggr) \,,
\end{align}
\begin{align}
I^P_1(\rho,Q)
&=\frac{9}{4 k_{\rm F}^3} \int^{k_\text{F}}_0 \int^1_{-1}
\frac{p^4(3\cos^2\theta-1)}{m_\pi^2+p^2+\frac{Q^2}{4}-p Q \cos\theta} \, dp \, d\cos\theta \,,
\nonumber \\
I^P_2(\rho,Q)
&=\frac{9}{2 k_{\rm F}^3}\int^{k_\text{F}}_0 \int^1_{-1}
\frac{p^3 Q \cos\theta}{m_\pi^2+p^2+\frac{Q^2}{4}-p Q \cos\theta} \, dp \, 
d\cos\theta \,, \nonumber \\
I^P_4(\rho,Q)
&=\frac{9}{2 k_{\rm F}^3} \int^{k_\text{F}}_0 \int^1_{-1}
\frac{p^2 Q^2}{m_\pi^2+p^2+\frac{Q^2}{4}-p Q \cos\theta} \, dp \, 
d\cos\theta \,, \nonumber \\[2mm]
&I^P_1(\rho,Q) - I^P_2(\rho,Q) + \frac{1}{4} \, I^P_4(\rho,Q) = I^P(\rho,Q) \,,
\end{align} 
\begin{align}
&I_{c_6}(\rho,Q)
=\frac{9}{2 k_{\rm F}^3} \int^{k_\text{F}}_0 \int^1_{-1}
\frac{\left(p^3Q\cos\theta-\frac{p^2Q^2}{2}\right)}{
m_\pi^2+p^2+\frac{Q^2}{4}-p Q \cos\theta} \, dp \, d\cos\theta 
\nonumber \\[1mm]
&=-\frac{9}{128 k_{\rm F}^3Q} \Biggl( 
\Bigl[ 32k_{\rm F}^3Q+32k_{\rm F}m_{\pi}^2Q+8k_{\rm F}Q^3 \Bigr]
+ \Bigl[ 16(k_{\rm F}^2+m_{\pi}^2)^2+8(m_{\pi}^2-k_{\rm F}^2)Q^2+Q^4
\Bigr]
 \log\biggl[\frac{4m_{\pi}^2+(2k_{\rm F}-Q)^2}{4m_{\pi}^2+(2k_{\rm F}+Q)^2}
\biggr] \Biggr) \,.
\end{align}

\section{Derivation of the structure factor $S_A(p)$}
\label{SFderivation}

We start from the Lagrangian density for spin-dependent WIMP-nucleus 
scattering Eq.~\eqref{L}. WIMPs are expected to be nonrelativistic
with velocities of the order $v/c \sim 10^{-3}$, so the time components
of the currents can be neglected. Evaluating the Lagrangian density
between initial and final states leads to
\begin{equation}
\bra{f} {\mathcal L}^{\rm SD}_\chi \ket{i} = -\frac{G_F}{\sqrt{2}}
\int d^3{\bf r} \, e^{-i{\bf p} \cdot {\bf r}} \, \overline{\chi}_f 
{\bm \gamma} \gamma_5 \chi_i \, {\bf J}^A_{fi}({\bf r}) \,,
\end{equation}
where $e^{-i{\bf p} \cdot {\bf r}} \, \overline{\chi}_f {\bm \gamma}
\gamma^5 \chi_i =\bra{\chi_f}{\bf j}({\bf r}) \ket{\chi_i}$ represents
the matrix element of the leptonic current of the WIMP and ${\bf
J}^A_{fi}({\bf r})$ that of the hadronic current.

We can expand the leptonic current in terms of spherical unit 
vectors~\cite{Walecka}:
\begin{equation}
\overline{\chi}_f {\bm \gamma} \gamma^5 \chi_i \, e^{-i{\bf p} \cdot {\bf r}}
= {\bf l} \, e^{-i{\bf p} \cdot {\bf r}} = \sum_{\lambda=0,\pm 1} l_\lambda
\, {\bf e}^\dagger_\lambda \, e^{-i{\bf p} \cdot {\bf r}} \,,
\label{leptonic}
\end{equation}
with spherical unit vectors with a $z$-axis in the direction of ${\bf p}$ 
\begin{align}
{\bf e}_{\pm1} &\equiv \mp \frac{1}{\sqrt{2}} ({\bf e}_{p1} \pm i 
{\bf e}_{p2}) &{\bf e}_0 &\equiv \frac{\bf p}{|{\bf p}|} \,, \\
l_{\pm 1} &= \mp \frac{1}{\sqrt{2}} (l_1\pm il_2) &l_{\lambda=0} &\equiv l_3 \,.
\end{align}
We can also expand the product ${\bf e}^\dagger_\lambda \,
e^{-i{\bf p} \cdot {\bf r}}$ in Eq.~\eqref{leptonic} in a multipole
expansion~\cite{Walecka}. This leads to
\begin{align}
&\bra{f}\!{\mathcal L}^{\rm SD}_\chi\!\ket{i}
\!=\! -\frac{G_F}{\sqrt{2}}\!\!\bra{J_f M_f}\!\!\Biggl( \sum_{L \geqslant 0}
\!\!\sqrt{4\pi (2L+1)}(-i)^L l_3 \, {\mathcal L}_{L0}^{5}(p)
\!-\!\!\sum_{L\geqslant 1}\!\!\sqrt{2\pi(2L+1)} (-i)^L
\!\!\sum_{\lambda=\pm 1}\!\!l_\lambda \!\!\left[
{\mathcal T}_{L-\lambda}^{\mathrm{el}5}(p)\!+\!\lambda
{\mathcal T}_{L-\lambda}^{\mathrm{mag}5}\!(p)\right]\!\!\Biggr)\!\!\ket{J_i M_i}\!,
\label{expansion}
\end{align}
where $\ket{J_i M_i}$, $\ket{J_f M_f}$ denote the initial and final 
states of the nucleus, $p=|{\bf p}|$. The electric longitudinal,
electric transverse, and magnetic transverse multipole operators
are defined by~\cite{Walecka}
\begin{align}
{\mathcal L}_{LM}^5(p) &= \frac{i}{p} \int d^3{\bf r} \, \Bigl[{\bm \nabla}
\bigl[j_L(pr)Y_{LM}(\Omega_r)\bigr]\Bigr] \cdot {\bf J}^A({\bf r}) \,, 
\label{multi1} \\
{\mathcal T}_{LM}^{\mathrm{el}5}(p) &= \frac{1}{p} \int d^3{\bf r} \,
\bigl[{\bm \nabla} \times j_L(pr){\bf Y}^M_{LL1}(\Omega_r)\bigr] 
\cdot {\bf J}^A({\bf r}) \,, \\
{\mathcal T}_{LM}^{\mathrm{mag}5}(p) &= \int d^3{\bf r} \, 
\bigl[j_L(pr){\bf Y}^M_{LL1}(\Omega_r)\bigr] \cdot {\bf J}^A({\bf r}) \,,
\label{multi3}
\end{align}
with spherical Bessel function $j_L(pr)$. The vector spherical 
harmonics are given by
\begin{equation}
{\bf Y}^M_{LL'1}(\Omega_r) = \sum_{m\lambda}\braket{L'm1\lambda|L'1LM}
Y_{L'm}(\Omega_r) \, {\bf e}_\lambda \,.
\end{equation}
Since ${\bf J}^A({\bf r})=\sum^A_{i=1}{\bf J}^A_i({\bf r})
\delta({\bf r}-{\bf r}_i)$, the multipole operators can be written
as a sum of one-body operators:
\begin{align}
{\mathcal L}_{LM}^5(p) &= \frac{i}{p} \sum^A_{i=1} 
\Bigl[{\bm \nabla}\bigl[j_L(pr_i)Y_{LM}({\bf r}_i)\bigr]\Bigr] 
\cdot {\bf J}^A_i({\bf r}_i) \nonumber \\
&= \frac{i}{\sqrt{2L+1}} \sum^A_{i=1} \Bigl[\sqrt{L+1} 
j_{L+1}(pr_i){\bf Y}^M_{L(L+1)1}({\bf r}_i)
+\sqrt{L} \, j_{L-1}(pr_i){\bf Y}^M_{L(L-1)1}({\bf r}_i)\Bigr]
\cdot {\bf J}^A_i({\bf r}_i) \,, \\
{\mathcal T}_{LM}^{\mathrm{el}5}(p) &= \frac{1}{p}\sum^A_{i=1}
\bigl[{\bm \nabla}\times j_L(pr_i){\bf Y}^M_{LL1}({\bf r}_i)\bigr]
\cdot {\bf J}^A_i({\bf r}_i) \nonumber \\
&= \frac{i}{\sqrt{2L+1}}\sum^A_{i=1}\Bigl[\sqrt{L+1} \, j_{L-1}(pr_i)
{\bf Y}^M_{L(L-1)1}({\bf r}_i)
-\sqrt{L} \, j_{L+1}(pr_i){\bf Y}^M_{L(L+1)1}({\bf r}_i)\Bigr]
\cdot {\bf J}^A_i({\bf r}_i) \,, \\
{\mathcal T}_{LM}^{\mathrm{mag}5}(p) &= \sum^A_{i=1}j_L(pr_i)
{\bf Y}^M_{LL1}({\bf r}_i)\cdot {\bf J}^A_i({\bf r}_i) \,.
\end{align}

The structure factor $S_A(p)$ is obtained from $\bigl| \bra{f}{
\mathcal L}^{\rm SD}_\chi\ket{i} \bigr|^2$ by summing over the final
neutralino spin and over the nucleus final-state angular momentum
projections, and by averaging over the initial configurations.  It is
thus useful to work with reduced matrix elements that do not depend on
projection numbers:
\begin{align}
\bra{J_f M_f}{O}_{LM}\ket{J_iM_i} %\\
=(-1)^{J_f-M_f}\left(
\begin{array}{ccc} J_f&L&J_i \\
-M_f&M&M_i \end{array} \right)
\bra{ J_f}\!|{O}_L|\!\ket{ J_i} \,,
\end{align}
with $3j$ coefficients and where $O$ is a tensor operator of rank $L$.
This gives for the sum and average~\cite{Walecka}
\begin{align}
\hspace{-0.08cm}
\frac{1}{2(2J_i+1)}\!\sum_{s_f,s_i}\! \sum_{M_f,M_i}\!\!
\bigl|\!\bra{f}\!{\mathcal L}^{\rm SD}_\chi\!\ket{i}\!\bigr|^2 
&\!=\! \frac{G_F^2}{4}\frac{1}{(2J_i+1)}\!\sum_{s_f,s_i}\! \biggl(
\!\sum_{L \geqslant 0}\! 4\pi l_3 l^*_3 \bigl|\!\bra{J_f}\!|
{\mathcal L}_L^5|\!\ket{J_i}\!\bigr|^2
\!+\!\sum_{\lambda=\pm 1}\! l_\lambda l^*_\lambda \!\sum_{L \geqslant 1}\!
2\pi \bigl|\!\bra{J_f}\!|{\mathcal T}_L^{\mathrm{el}5}
\!+\!\lambda{\mathcal T}_L^{\mathrm{mag}5}|\!\ket{J_i}\!\bigr|^2 \!\biggr) \nonumber \\
&\!= \frac{G_F^2}{4} \frac{4\pi}{(2J_i+1)} \sum_{s_f,s_i}
\biggl( \sum_{L \geqslant 0} l_3l_3^* \bigl|
\bra{J_f}\!|{\mathcal L}_L^5|\!\ket{J_i}\bigr|^2
\!+\!\sum_{L \geqslant 1}\biggl[\frac{1}{2}({\bf l}\cdot{\bf l}^*-l_3l_3^*)
\Bigl(\bigl|\bra{ J_f}\!|{\mathcal T}_L^{\mathrm{el}5}|\!\ket{J_i}\bigr|^2 \nonumber \\
&\quad+\bigl|\bra{J_f}\!|{\mathcal T}_L^{\mathrm{mag}5}|\!\ket{J_i}\bigr|^2
\Bigr)\!-\!\frac{i}{2}({\bf l}\times{\bf l}^*)_3
\Bigl(2 \, \text{Re}\bra{ J_f}\!|{\mathcal T}_L^{\mathrm{el}5}|\!\ket{J_i}
\bra{J_f}\!|{\mathcal T}_L^{\mathrm{mag}5}|\!\ket{J_i}^* \Bigr) \biggr]
\biggr),
\end{align}
where we have assumed that the neutralino spin is $1/2$, and the cross
terms vanish due to the orthogonalization properties of the $3j$
coefficients. For the sum over neutralino spin projections one has
for $\mu,\nu=1,2,3$
\begin{align}
-\sum_{s_i,s_f}l_\mu l_\nu^*
&= \sum_{s_i,s_f}\overline{\chi}^{s_f}(p_f)\gamma^\mu \gamma^5 \chi^{s_i}(p_i)
\, \overline{\chi}^{s_i}(p_i)\gamma^5\gamma^\nu\chi^{s_f}(p_f) \,, \nonumber \\
&= \sum_{s_i,s_f} \bigl(\chi^{s_f}_\delta(p_f)\overline{\chi}^{s_f}_\alpha(p_f)
(\gamma^\mu\gamma^5)_{\alpha\beta} \chi^{s_i}_\beta(p_i)
\overline{\chi}^{s_i}_\gamma(p_i)(\gamma^5\gamma^\nu)_{\gamma\delta}\bigr) ,
\nonumber \\
&=\frac{1}{4}\left[2\text{Tr}(\gamma^0\gamma^\mu \gamma^5 \gamma^5 \gamma^\nu)
+2\text{Tr}(\gamma^\mu \gamma^5 \gamma^5 \gamma^\nu)\right]
=\frac{1}{2}\text{Tr}(\gamma^\mu \gamma^5 \gamma^5 \gamma^\nu) = -2\delta^{\mu\nu} \,,
\end{align}
which follows from the completeness relation
\begin{equation}
\sum_s \chi_\alpha^{s}(p) \overline{\chi}_\beta^{s}(p) = 
\Bigl(\frac{p_\mu \gamma^\mu+m}{2E_p}\Bigr)_{\alpha \beta}
\approx \frac{1}{2}\left(\gamma^0+\mathbb{1}\right)_{\alpha\beta} \,,
\end{equation}
valid for nonrelativistic WIMPs.
Combined, this gives the final result:
\begin{align}
\frac{1}{2(2J_i+1)}\sum_{s_f,s_i}\sum_{M_f,M_i}
\bigl|\bra{f}{\mathcal L}^{\rm SD}_\chi\ket{i}\bigr|^2
=\frac{G_F^2}{2}\frac{4\pi}{(2J_i+1)}
\biggl[ \sum_{L\geqslant 0} 
\bigl|\bra{J_f}\!|{\mathcal L}_L^5|\!\ket{J_i}\bigr|^2
+ \sum_{L \geqslant 1} \Bigl(
\bigl|\bra{ J_f}\!|{\mathcal T}_L^{\mathrm{el}5}|\!\ket{J_i}\bigr|^2
+\bigl|\bra{J_f}\!|{\mathcal T}_L^{\mathrm{mag}5}|\!\ket{ J_i}\bigr|^2
\Bigr) \biggr].
\end{align}

The specific form of the multipoles depends on the form of the WIMP
currents ${\bf J}^A_i({\bf r})$. They contain either axial-vector terms 
[${\bm \sigma_i}$] or pseudo-scalar ones [$({\bf p}\cdot{\bm \sigma_i})
{\bf p}$]. For axial-vector currents, the response will be proportional
to the following operator
\begin{align}
M^M_{L,L'}(p{\bf r}_i) &= j_{L'}(pr_i){\bf Y}^M_{LL1}({\bf r}_i)
\cdot{\bm \sigma_i} \,, \nonumber \\[1mm]
&= j_{L'}(pr_i)\sum_{m\lambda}\braket{L'm1\lambda|L'1LM}
Y_{L'm}({\bf r}_i) \, \sigma_i^{1\lambda}= j_{L'}(pr_i) \bigl[Y_{L'}({\bf r}_i){\bm \sigma}_i\bigr]^L \,.
\end{align}
Pseudo-Scalar currents, which are proportional to the momentum transfer
${\bf p}$, only contribute to the longitudinal multipoles (see
Eq.~\eqref{expansion}). Moreover, in these we can replace $({\bf p}
\cdot{\bm \sigma}_i){\bf p}$ by $p^2 {\bm \sigma}_i$, because of
\begin{equation}
({\bf p}\cdot{\bm \sigma_i}){\bf p} = 
p^2{\bm \sigma_i}+{\bf p}\times({\bf p}\times{\bm \sigma_i}) \,,
\end{equation}
and the second term is perpendicular to ${\bf p}$, so it vanishes for
the longitudinal multipoles. As a result, pseudo-scalar currents can
also be expressed in terms of $M^M_{L,L'}(p{\bf r}_i)$.

In summary, including chiral 2b currents at the normal-ordered
one-body level in Eq.~\eqref{J3}, we have for the multipoles
\begin{align}
{\mathcal L}_{L}^5(p) =& \frac{i}{\sqrt{2L+1}}
\sum^A_{i=1} \frac{1}{2}\biggl[a_0+a_1 \tau^3_i
\Bigl(1+\delta a_1(p)-\frac{2g_{\pi pn}F_\pi p^2}{2mg_A
(p^2+m^2_\pi)}+\delta a_1^P(p)\Bigr)\biggr] \nonumber \\[1mm]
&\times \Bigl[\sqrt{L+1} M_{L,L+1}(p{\bf r}_i)+
\sqrt{L} M_{L,L-1}(p{\bf r}_i)\Bigr] \,, \\[2mm]
{\mathcal T}_{L}^{\mathrm{el}5}(p) =& \frac{i}{\sqrt{2L+1}}
\sum^A_{i=1}\frac{1}{2} \biggl[a_0+a_1\tau^3_i
\Bigl(1-2\frac{p^2}{\Lambda^2_A}+\delta a_1(p)\Bigr)\biggr]\Bigl[-\sqrt{L}M_{L,L+1}(p{\bf r}_i)
+\sqrt{L+1} M_{L,L-1}(p{\bf r}_i)\Bigr], \\[2mm]
%\end{align}
%\begin{align}
{\mathcal T}_{L}^{\mathrm{mag}5}(p)=&\sum^A_{i=1}\frac{1}{2}\biggl[a_0+a_1\tau^3_i
\Bigl(1-2\frac{p^2}{\Lambda^2_A}+\delta a_1(p)\Bigr)\biggr] 
M_{L,L}(p{\bf r}_i) \,.
\end{align}
Note that the $p^2/\Lambda_A^2$ terms cancel in the longitudinal
response and only contribute to the transverse multipoles.

\section{Reduced matrix elements of $M_{L,L'}(p {\bf r}_i)$}
\label{mes}

To calculate the structure factor, we need the matrix elements
of the one-body operator $M_{L,L'}(p {\bf r}_i) = j_{L'}(p r_i)\times$
$[Y_{L'}(\hat{\bf r}_i) \, {\bm \sigma}_i]^L$ between the
single-particle states of the many-body basis used for the description
of the nuclear states. The reduced matrix elements can be obtained as
a function of $3j$ and $9j$ symbols and matrix elements of the
spherical Bessel functions $j_{L'}$,
\begin{align}
&\bra{n'l'\frac{1}{2}j'}\!|M_{L,L'}(p{\bf r}_i)|\!\ket{nl\frac{1}{2}j}
\nonumber \\
&= \sum_{n''l''} \bra{n'l'}\!|j_{L'}Y_{L'}|\!\ket{n''l''}
\bra{n''l''\frac{1}{2}}\!|{\bm \sigma}_i|\!\ket{nl\frac{1}{2}}
\bigl[(2j+1)(2j'+1)(2L+1)\bigr]^\frac{1}{2}\left\lbrace
\begin{array}{ccc}
l' & l & L' \\
\frac{1}{2} & \frac{1}{2} & 1 \\
j' & j & L
\end{array}\right\rbrace \nonumber \\
&=\bra{n'l'}j_{L'}\ket{nl} \bra{n'l'}\!|Y_{L'}|\!\ket{nl}
\bra{\frac{1}{2}}\!|{\bm \sigma}_i|\!\ket{\frac{1}{2}}
\bigl[(2j+1)(2j'+1)(2L+1)\bigr]^\frac{1}{2}\left\lbrace
\begin{array}{ccc}
l' & l & L' \\
\frac{1}{2} & \frac{1}{2} & 1 \\
j' & j & L
\end{array}\right\rbrace \nonumber \\
&=\bra{n'l'}j_{L'}(pr_i)\ket{nl} (-1)^{l'}
\sqrt{\frac{6}{4\pi}} \, \bigl[(2l'+1)(2l+1)(2j'+1)(2j+1)\bigr]^{\frac{1}{2}}
\bigl[(2L'+1)(2L+1)\bigr]^\frac{1}{2} \nonumber \\
&\quad\times \left(
\begin{array}{ccc}
l' & L' & l \\
0 & 0 & 0
\end{array}\right)			
\left\lbrace
\begin{array}{ccc}
l' & l & L' \\
\frac{1}{2} & \frac{1}{2} & 1 \\
j' & j & L
\end{array}\right\rbrace \,.
\end{align}

\section{Fits of the structure factors}
\label{fits}

In Tables~\ref{xenon_fit}-\ref{si_fit} we give fits for the isoscalar/isovector
and ``neutron-only''/``proton-only'' decompositions of the structure factors of all isotopes studied in this work.
Results are given including 1b+2b currents.

\onecolumngrid

\begin{table}[h]
%\centering {\bf Supplemental Material}
\caption{Fits to the isoscalar/isovector structure factors $S_{00}$, 
$S_{11}$ and $S_{01}$ as well as ``proton-only" and ``neutron-only"
structure factors $S_{p}$ and $S_{n}$ for spin-dependent WIMP elastic
scattering off $^{129}$Xe and $^{131}$Xe nuclei,
including 1b and 2b currents as in Fig.~6. The upper and lower limits
from the theoretical error band were used for the fit.  The fitting
function of the dimensionless variable $u = p^2 b^2/2$ is $S_{ij}(u) =
e^{-u} \sum_{n=0}^9 c_{ij,n} u^n$. The rows give the coefficients
$c_{ij,n}$ of the $u^n$ terms in the polynomial.\label{xenon_fit}}
\begin{center}
\begin{tabular*}{0.773\textwidth}{c||c|c|c|c|c}
\hline
\multicolumn{6}{c}{$^{129}$Xe} \\
\multicolumn{6}{c}{$u=p^2b^2/2 \,, \: b=2.2853 \, {\rm fm}$} \\
\hline
$e^{-u}\times$ & $S_{00}$ & $S_{11}$ (1b+2b min) & $S_{11}$ (1b+2b max) & 
$S_{01}$ (1b+2b min) & $S_{01}$ (1b+2b max) \\
\hline
$1$ & $0.0547144$ & $0.0221559$ & $0.0357742$ & $-0.0885644$ & $-0.0696691 $\\
$u$ & $-0.146407$ & $-0.0656100$ & $-0.107895$ & $0.254049$ & $0.197380 $\\
$u^2$ & $0.180603$ & $0.0863920$ & $0.145055$ & $-0.332322$ & $-0.254839 $\\
$u^3$ & $-0.125526$ & $-0.0631729$ & $-0.108549$ & $0.244981$ & $0.185896 $\\
$u^4$ & $0.0521484$ & $0.0278792$ & $0.0490401$ & $-0.109298$ & $-0.0825294 $\\
$u^5$ & $-0.0126363$ & $-0.00756661$ & $-0.0136169$ & $0.0296705$ & $0.0224322 $\\
$u^6$ & $0.00176284$ & $0.00126767$ & $0.00233283$ & $-0.00492657$ & $-0.00375109 $\\
$u^7$ & $-1.32501 \times 10^{-4}$ & $-1.27755 \times 10^{-4}$ & $-2.39926 \times 10^{-4}$ & $4.88467 \times 10^{-4}$ & $3.77179 \times 10^{-4} $\\
$u^8$ & $4.23423 \times 10^{-6}$ & $7.10322 \times 10^{-6}$ & $1.35553 \times 10^{-5}$ & $-2.65022 \times 10^{-5}$ & $-2.09510 \times 10^{-5} $\\
$u^9$ & $-1.68052 \times 10^{-9}$ & $-1.67272 \times 10^{-7}$ & $-3.21404 \times 10^{-7}$ & $5.98909 \times 10^{-7}$ & $4.92362 \times 10^{-7} $\\
\hline
$e^{-u}\times$ & & $S_{p}$ (1b+2b min) & $S_{p}$ (1b+2b max)& 
$S_{n}$ (1b+2b min) & $S_{n}$ (1b+2b max) \\
\hline
$1$&&$ 0.00196369$ & $0.00715281$ & $0.146535$ & $0.179056 $\\
$u$&&$ -0.00119154$ & $-0.0134790$ & $-0.409290$ & $-0.508334 $\\
$u^2$&&$ -0.00324210$ & $0.00788823$ & $0.521423$ & $0.657560 $\\
$u^3$&&$ 0.00622602$ & $0.00311153$ & $-0.374011$ & $-0.477988 $\\
$u^4$&&$ -0.00496653$ & $-0.00653771$ & $0.162155$ & $0.209437 $\\
$u^5$&&$ 0.00224469$ & $0.00375478$ & $-0.0424842$ & $-0.0554186 $\\
$u^6$&&$ -5.74412 \times 10^{-4}$ & $-0.00105558$ & $0.00674911$ & $0.00889251 $\\
$u^7$&&$ 8.31313 \times 10^{-5}$ & $1.59440 \times 10^{-4}$ & $-6.33434 \times 10^{-4}$ & $-8.42977 \times 10^{-4} $\\
$u^8$&&$ -6.41114 \times 10^{-6}$ & $-1.25055 \times 10^{-5}$ & $3.20266 \times 10^{-5}$ & $4.30517 \times 10^{-5} $\\
$u^9$&&$ 2.07744 \times 10^{-7}$ & $4.04987 \times 10^{-7}$ & $-6.54245 \times 10^{-7}$ & $-8.88774 \times 10^{-7} $\\
\hline
\hline
\multicolumn{6}{c}{$^{131}$Xe} \\
\multicolumn{6}{c}{$u=p^2b^2/2 \,, \: b=2.2905 \, {\rm fm}$} \\
\hline 
$e^{-u}\times$ & $S_{00}$ & $S_{11}$ (1b+2b min) & $S_{11}$ (1b+2b max) &
$S_{01}$ (1b+2b min) & $S_{01}$ (1b+2b max) \\
\hline
$1$ & $0.0417857$ & $0.0167361$ & $0.0271052$ & $-0.0675438$ & $-0.0529487 $\\
$u$ & $-0.111132$ & $-0.0472853$ & $-0.0812985$ & $0.195710$ & $0.146987 $\\
$u^2$ & $0.171306$ & $0.0684924$ & $0.122960$ & $-0.306688$ & $-0.225003 $\\
$u^3$ & $-0.132481$ & $-0.0514413$ & $-0.0940491$ & $0.243678$ & $0.179499 $\\
$u^4$ & $0.0630161$ & $0.0237858$ & $0.0439746$ & $-0.118395$ & $-0.0888278 $\\
$u^5$ & $-0.0177684$ & $-0.00692778$ & $-0.0128013$ & $0.0351428$ & $0.0271514 $\\
$u^6$ & $0.00282192$ & $0.00124370$ & $0.00227407$ & $-0.00622577$ & $-0.00499280 $\\
$u^7$ & $-2.32247 \times 10^{-4}$ & $-1.31617 \times 10^{-4}$ & $-2.35642 \times 10^{-4}$ & $6.31685 \times 10^{-4}$ & $5.31148 \times 10^{-4} $\\
$u^8$ & $7.81471 \times 10^{-6}$ & $7.46669 \times 10^{-6}$ & $1.28691 \times 10^{-5}$ & $-3.33272 \times 10^{-5}$ & $-2.99162 \times 10^{-5} $\\
$u^9$ & $1.25984 \times 10^{-9}$ & $-1.73484 \times 10^{-7}$ & $-2.77011 \times 10^{-7}$ & $6.82500 \times 10^{-7}$ & $6.81902 \times 10^{-7} $\\
\hline
$e^{-u}\times$ & & $S_{p}$ (1b+2b min) & $S_{p}$ (1b+2b max)& 
$S_{n}$ (1b+2b min) & $S_{n}$ (1b+2b max) \\
\hline
$1$&&$ 0.00159352$ & $0.00529643$ & $0.111627$ & $0.136735 $\\
$u$&&$ -0.00207344$ & $-0.00528808$ & $-0.308602$ & $-0.393930 $\\
$u^2$&&$ 0.00567412$ & $-0.00627452$ & $0.474842$ & $0.617924 $\\
$u^3$&&$ -0.00605643$ & $0.0227436$ & $-0.375201$ & $-0.488443 $\\
$u^4$&&$ 0.00337794$ & $-0.0192229$ & $0.182382$ & $0.234645 $\\
$u^5$&&$ -6.88135 \times 10^{-4}$ & $0.00844826$ & $-0.0539711$ & $-0.0681357 $\\
$u^6$&&$ -3.42717 \times 10^{-5}$ & $-0.00212755$ & $0.00944180$ & $0.0116393 $\\
$u^7$&&$ 3.13222 \times 10^{-5}$ & $3.03972 \times 10^{-4}$ & $-9.34456 \times 10^{-4}$ & $-0.00111487 $\\
$u^8$&&$ -4.02617 \times 10^{-6}$ & $-2.27893 \times 10^{-5}$ & $4.73386 \times 10^{-5}$ & $5.34878 \times 10^{-5} $\\
$u^9$&&$ 1.72711 \times 10^{-7}$ & $7.05661 \times 10^{-7}$ & $-9.01514 \times 10^{-7}$ & $-9.03594 \times 10^{-7} $\\
\hline
\end{tabular*}
\end{center}
\end{table}

\newpage

\onecolumngrid

\begin{table}[t]
\caption{Fits to the isoscalar/isovector structure factors $S_{00}$, 
$S_{11}$ and $S_{01}$ as well as ``proton-only" and ``neutron-only"
structure factors $S_{p}$ and $S_{n}$ for spin-dependent WIMP elastic
scattering off $^{73}$Ge (Int. 2) and $^{127}$I nuclei,
including 1b and 2b currents as in Fig.~10 and Fig.~11. The upper and lower limits
from the theoretical error band were used for the fit.  The fitting
function of the dimensionless variable $u = p^2 b^2/2$ is $S_{ij}(u) =
e^{-u} \sum_{n=0}^9 c_{ij,n} u^n$. The rows give the coefficients
$c_{ij,n}$ of the $u^n$ terms in the polynomial.\label{gei_fit}}
\begin{center}
\begin{tabular*}{0.773\textwidth}{c||c|c|c|c|c}
\hline
\multicolumn{6}{c}{$^{73}$Ge (Int.2)} \\
\multicolumn{6}{c}{$u=p^2b^2/2 \,, \: b=2.1058 \, {\rm fm}$} \\
\hline
$e^{-u}\times$ & $S_{00}$ & $S_{11}$ (1b+2b min) & $S_{11}$ (1b+2b max) & 
$S_{01}$ (1b+2b min) & $S_{01}$ (1b+2b max) \\
\hline
$1$ & $0.215608$ & $0.0743728$ & $0.120045$ & $-0.321836$ & $-0.253289 $\\
$u$ & $-0.578786$ & $-0.233814$ & $-0.384157$ & $0.950136$ & $0.739394 $\\
$u^2$ & $0.698020$ & $0.341725$ & $0.559728$ & $-1.27413$ & $-0.993188 $\\
$u^3$ & $-0.372000$ & $-0.259024$ & $-0.415686$ & $0.831035$ & $0.659953 $\\
$u^4$ & $0.107576$ & $0.121206$ & $0.188412$ & $-0.323769$ & $-0.269522 $\\
$u^5$ & $-0.0182408$ & $-0.0371226$ & $-0.0568025$ & $0.0831244$ & $0.0745897 $\\
$u^6$ & $0.00217108$ & $0.00741080$ & $0.0120204$ & $-0.0151542$ & $-0.0144162 $\\
$u^7$ & $-2.07981 \times 10^{-4}$ & $-9.02610 \times 10^{-4}$ & $-0.00175855$ & $0.00193259$ & $0.00181542 $\\
$u^8$ & $1.65907 \times 10^{-5}$ & $5.81933 \times 10^{-5}$ & $1.59975 \times 10^{-4}$ & $-1.55025 \times 10^{-4}$ & $-1.29365 \times 10^{-4} $\\
$u^9$ & $-5.95664 \times 10^{-7}$ & $-1.38557 \times 10^{-6}$ & $-6.66472 \times 10^{-6}$ & $5.68777 \times 10^{-6}$ & $3.77020 \times 10^{-6} $\\
\hline
$e^{-u}\times$ & & $S_{p}$ (1b+2b min) & $S_{p}$ (1b+2b max)& 
$S_{n}$ (1b+2b min) & $S_{n}$ (1b+2b max) \\
\hline
$1$&&$ 0.0138433$ & $0.0366954$ & $0.543270$ & $0.657509 $\\
$u$&&$ -0.0138982$ & $-0.0733258$ & $-1.55198$ & $-1.91400 $\\
$u^2$&&$ -0.00961825$ & $0.0471313$ & $2.03269$ & $2.53820 $\\
$u^3$&&$ 0.0275620$ & $0.0281229$ & $-1.28990$ & $-1.63488 $\\
$u^4$&&$ -0.0101577$ & $-0.0405538$ & $0.496419$ & $0.639763 $\\
$u^5$&&$ -0.00235492$ & $0.0196085$ & $-0.128347$ & $-0.171656 $\\
$u^6$&&$ 0.00246030$ & $-0.00515247$ & $0.0232676$ & $0.0345442 $\\
$u^7$&&$ -6.53041 \times 10^{-4}$ & $8.06626 \times 10^{-4}$ & $-0.00274482$ & $-0.00504185 $\\
$u^8$&&$ 7.84526 \times 10^{-5}$ & $-6.95571 \times 10^{-5}$ & $1.81026 \times 10^{-4}$ & $4.64828 \times 10^{-4} $\\
$u^9$&&$ -3.61078 \times 10^{-6}$ & $2.63102 \times 10^{-6}$ & $-4.56383 \times 10^{-6}$ & $-1.93402 \times 10^{-5} $\\
\hline
\hline
\multicolumn{6}{c}{$^{127}$I} \\
\multicolumn{6}{c}{$u=p^2b^2/2 \,, \: b=2.2801 \, {\rm fm}$} \\
\hline 
$e^{-u}\times$ & $S_{00}$ & $S_{11}$ (1b+2b min) & $S_{11}$ (1b+2b max) &
$S_{01}$ (1b+2b min) & $S_{01}$ (1b+2b max) \\
\hline
$1$ & $0.0928480$ & $0.0297755$ & $0.0480576$ & $0.105154$ & $0.133610 $\\
$u$ & $-0.252496$ & $-0.0904582$ & $-0.148155$ & $-0.302437$ & $-0.388379 $\\
$u^2$ & $0.351982$ & $0.145234$ & $0.234436$ & $0.452142$ & $0.579490 $\\
$u^3$ & $-0.260427$ & $-0.132020$ & $-0.205618$ & $-0.371193$ & $-0.471030 $\\
$u^4$ & $0.118280$ & $0.0769978$ & $0.113448$ & $0.192342$ & $0.238903 $\\
$u^5$ & $-0.0319614$ & $-0.0290350$ & $-0.0396327$ & $-0.0631442$ & $-0.0751672 $\\
$u^6$ & $0.00492618$ & $0.00701812$ & $0.00870215$ & $0.0130940$ & $0.0144759 $\\
$u^7$ & $-4.06546 \times 10^{-4}$ & $-0.00105740$ & $-0.00116942$ & $-0.00169645$ & $-0.00166889 $\\
$u^8$ & $1.55818 \times 10^{-5}$ & $9.11013 \times 10^{-5}$ & $8.85742 \times 10^{-5}$ & $1.28905 \times 10^{-4}$ & $1.07845 \times 10^{-4} $\\
$u^9$ & $-1.64934 \times 10^{-7}$ & $-3.44003 \times 10^{-6}$ & $-2.91582 \times 10^{-6}$ & $-4.47150 \times 10^{-6}$ & $-3.09335 \times 10^{-6} $\\
\hline
$e^{-u}\times$ & & $S_{p}$ (1b+2b min) & $S_{p}$ (1b+2b max)& 
$S_{n}$ (1b+2b min) & $S_{n}$ (1b+2b max) \\
\hline
$1$&&$ 0.227779$ & $0.274511$ & $0.00729876$ & $0.0174634 $\\
$u$&&$ -0.645502$ & $-0.788708$ & $-0.0124606$ & $-0.0401552 $\\
$u^2$&&$ 0.950398$ & $1.16333$ & $0.00820860$ & $0.0429504 $\\
$u^3$&&$ -0.766815$ & $-0.929643$ & $0.00187492$ & $-0.0171587 $\\
$u^4$&&$ 0.391958$ & $0.460285$ & $-0.00353024$ & $-5.50598 \times 10^{-4} $\\
$u^5$&&$ -0.127209$ & $-0.138933$ & $0.00121496$ & $0.00367288 $\\
$u^6$&&$ 0.0262471$ & $0.0247388$ & $5.05292 \times 10^{-5}$ & $-0.00150561 $\\
$u^7$&&$ -0.00342824$ & $-0.00242940$ & $-1.09891 \times 10^{-4}$ & $2.73729 \times 10^{-4} $\\
$u^8$&&$ 2.66810 \times 10^{-4}$ & $1.08740 \times 10^{-4}$ & $2.14196 \times 10^{-5}$ & $-2.38605 \times 10^{-5} $\\
$u^9$&&$ -9.56532 \times 10^{-6}$ & $-8.75631 \times 10^{-7}$ & $-1.29204 \times 10^{-6}$ & $8.31918 \times 10^{-7} $\\
\hline
\end{tabular*}
\end{center}
\end{table}

\newpage

\onecolumngrid

\begin{table}[t]
\caption{Fits to the isoscalar/isovector structure factors $S_{00}$, 
$S_{11}$ and $S_{01}$ as well as ``proton-only" and ``neutron-only"
structure factors $S_{p}$ and $S_{n}$ for spin-dependent WIMP elastic
scattering off $^{19}$F,
including 1b and 2b currents as in Fig.~12. The upper and lower limits
from the theoretical error band were used for the fit.  The fitting
function of the dimensionless variable $u = p^2 b^2/2$ is $S_{ij}(u) =
e^{-u} \sum_{n=0}^{14} c_{ij,n} u^n$. The rows give the coefficients
$c_{ij,n}$ of the $u^n$ terms in the polynomial.\label{fluorine_fit}}
\begin{center}
\begin{tabular*}{0.773\textwidth}{c||c|c|c|c|c}
\hline
\multicolumn{6}{c}{$^{19}$F } \\
\multicolumn{6}{c}{$u=p^2b^2/2 \,, \: b=1.8032 \, {\rm fm}$} \\
\hline
$e^{-u}\times$ & $S_{00}$ & $S_{11}$ (1b+2b min) & $S_{11}$ (1b+2b max) & 
$S_{01}$ (1b+2b min) & $S_{01}$ (1b+2b max) \\
\hline
$1$ & $0.108058$ & $0.0505180$ & $0.0815382$ & $0.147769$ & $0.187748 $\\
$u$ & $-0.143789$ & $-0.102657$ & $-0.172679$ & $-0.248324$ & $-0.324839 $\\
$u^2$ & $0.0680848$ & $0.111644$ & $0.212269$ & $0.196804$ & $0.292189 $\\
$u^3$ & $4.07415 \times 10^{-4}$ & $-0.103800$ & $-0.228208$ & $-0.110517$ & $-0.243481 $\\
$u^4$ & $-0.0314817$ & $0.0920875$ & $0.213050$ & $0.0431978$ & $0.225724 $\\
$u^5$ & $0.0385933$ & $-0.0693892$ & $-0.153539$ & $0.00355133$ & $-0.187879 $\\
$u^6$ & $-0.0293716$ & $0.0406756$ & $0.0811970$ & $-0.0214773$ & $0.120370 $\\
$u^7$ & $0.0152264$ & $-0.0180247$ & $-0.0312282$ & $0.0171137$ & $-0.0567987 $\\
$u^8$ & $-0.00552655$ & $0.00597662$ & $0.00872716$ & $-0.00777410$ & $0.0195241 $\\
$u^9$ & $0.00141965$ & $-0.00146688$ & $-0.00176305$ & $0.00231495$ & $-0.00485435 $\\
$u^{10}$ & $-2.56989 \times 10^{-4}$ & $2.61654 \times 10^{-4}$ & $2.53666 \times 10^{-4}$ & $-4.67535 \times 10^{-4}$ & $8.61430 \times 10^{-4} $\\
$u^{11}$ & $3.20688 \times 10^{-5}$ & $-3.28624 \times 10^{-5}$ & $-2.52190 \times 10^{-5}$ & $6.36451 \times 10^{-5}$ & $-1.06203 \times 10^{-4} $\\
$u^{12}$ & $-2.62562 \times 10^{-6}$ & $2.74752 \times 10^{-6}$ & $1.63658 \times 10^{-6}$ & $-5.60211 \times 10^{-6}$ & $8.63415 \times 10^{-6} $\\
$u^{13}$ & $1.26950 \times 10^{-7}$ & $-1.36980 \times 10^{-7}$ & $-6.18772 \times 10^{-8}$ & $2.88239 \times 10^{-7}$ & $-4.15920 \times 10^{-7} $\\
$u^{14}$ & $-2.74719 \times 10^{-9}$ & $3.07589 \times 10^{-9}$ & $1.02158 \times 10^{-9}$ & $-6.58792 \times 10^{-9}$ & $8.98798 \times 10^{-9} $\\
\hline
$e^{-u}\times$ & & $S_{p}$ (1b+2b min) & $S_{p}$ (1b+2b max)& 
$S_{n}$ (1b+2b min) & $S_{n}$ (1b+2b max) \\
\hline
$1$&&$ 0.306344$ & $0.377350$ & $0.00186788$ & $0.0108048 $\\
$u$&&$ -0.494703$ & $-0.641645$ & $0.00680710$ & $0.00209733 $\\
$u^2$&&$ 0.375778$ & $0.575714$ & $0.00639787$ & $-0.0195694 $\\
$u^3$&&$ -0.210605$ & $-0.482204$ & $-0.0611310$ & $0.0180694 $\\
$u^4$&&$ 0.0963209$ & $0.426127$ & $0.114287$ & $-0.00732843 $\\
$u^5$&&$ -0.0171498$ & $-0.322095$ & $-0.118072$ & $-0.00123149 $\\
$u^6$&&$ -0.0189635$ & $0.185010$ & $0.0795624$ & $0.00434979 $\\
$u^7$&&$ 0.0194977$ & $-0.0786211$ & $-0.0371512$ & $-0.00349429 $\\
$u^8$&&$ -0.00944981$ & $0.0245769$ & $0.0123395$ & $0.00167052 $\\
$u^9$&&$ 0.00288142$ & $-0.00561387$ & $-0.00293887$ & $-5.31956 \times 10^{-4} $\\
$u^{10}$&&$ -5.87122 \times 10^{-4}$ & $9.23589 \times 10^{-4}$ & $4.98543 \times 10^{-4}$ & $1.15596 \times 10^{-4} $\\
$u^{11}$&&$ 8.01160 \times 10^{-5}$ & $-1.06384 \times 10^{-4}$ & $-5.88110 \times 10^{-5}$ & $-1.69465 \times 10^{-5} $\\
$u^{12}$&&$ -7.04748 \times 10^{-6}$ & $8.13277 \times 10^{-6}$ & $4.58527 \times 10^{-6}$ & $1.60329 \times 10^{-6} $\\
$u^{13}$&&$ 3.61875 \times 10^{-7}$ & $-3.70365 \times 10^{-7}$ & $-2.12430 \times 10^{-7}$ & $-8.83654 \times 10^{-8} $\\
$u^{14}$&&$ -8.24953 \times 10^{-9}$ & $7.60000 \times 10^{-9}$ & $4.42852 \times 10^{-9}$ & $2.15466 \times 10^{-9} $\\
\hline
\end{tabular*}
\end{center}
\end{table}

\newpage

\onecolumngrid

\begin{table}[t]
\caption{Fits to the isoscalar/isovector structure factors $S_{00}$, 
$S_{11}$ and $S_{01}$ as well as ``proton-only" and ``neutron-only"
structure factors $S_{p}$ and $S_{n}$ for spin-dependent WIMP elastic
scattering off $^{23}$Na and $^{27}$Al nuclei,
including 1b and 2b currents as in Fig.~13. The upper and lower limits
from the theoretical error band were used for the fit.  The fitting
function of the dimensionless variable $u = p^2 b^2/2$ is $S_{ij}(u) =
e^{-u} \sum_{n=0}^9 c_{ij,n} u^n$. The rows give the coefficients
$c_{ij,n}$ of the $u^n$ terms in the polynomial.\label{naal_fit}}
\begin{center}
\begin{tabular*}{0.773\textwidth}{c||c|c|c|c|c}
\hline
\multicolumn{6}{c}{$^{23}$Na} \\
\multicolumn{6}{c}{$u=p^2b^2/2 \,, \: b=1.8032 \, {\rm fm}$} \\
\hline
$e^{-u}\times$ & $S_{00}$ & $S_{11}$ (1b+2b min) & $S_{11}$ (1b+2b max) & 
$S_{01}$ (1b+2b min) & $S_{01}$ (1b+2b max) \\
\hline
$1$ & $0.0325305$ & $0.00973487$ & $0.0157138$ & $0.0356077$ & $0.0453141 $\\
$u$ & $-0.0433531$ & $-0.0185306$ & $-0.0312138$ & $-0.0582455$ & $-0.0772792 $\\
$u^2$ & $0.0319487$ & $0.0199627$ & $0.0351984$ & $0.0551609$ & $0.0769308 $\\
$u^3$ & $-0.00568858$ & $-0.00905267$ & $-0.0180647$ & $-0.0210939$ & $-0.0327180 $\\
$u^4$ & $2.67783 \times 10^{-4}$ & $0.00207003$ & $0.00580816$ & $0.00499454$ & $0.00946296 $\\
$u^5$ & $2.44643 \times 10^{-5}$ & $-2.28653 \times 10^{-4}$ & $-0.00122900$ & $-9.09266 \times 10^{-4}$ & $-0.00199807 $\\
$u^6$ & $-4.79620 \times 10^{-6}$ & $4.31460 \times 10^{-6}$ & $1.72086 \times 10^{-4}$ & $1.28051 \times 10^{-4}$ & $2.89585 \times 10^{-4} $\\
$u^7$ & $5.39846 \times 10^{-7}$ & $1.67535 \times 10^{-6}$ & $-1.52834 \times 10^{-5}$ & $-1.20016 \times 10^{-5}$ & $-2.59681 \times 10^{-5} $\\
$u^8$ & $-3.24691 \times 10^{-8}$ & $-1.67911 \times 10^{-7}$ & $7.73042 \times 10^{-7}$ & $6.29181 \times 10^{-7}$ & $1.25857 \times 10^{-6} $\\
$u^9$ & $8.09358 \times 10^{-10}$ & $5.14559 \times 10^{-9}$ & $-1.67756 \times 10^{-8}$ & $-1.39823 \times 10^{-8}$ & $-2.47908 \times 10^{-8} $\\
\hline
$e^{-u}\times$ & & $S_{p}$ (1b+2b min) & $S_{p}$ (1b+2b max)& 
$S_{n}$ (1b+2b min) & $S_{n}$ (1b+2b max) \\
\hline
$1$&&$ 0.0778747$ & $0.0935155$ & $0.00295139$ & $0.00674243 $\\
$u$&&$ -0.120203$ & $-0.151102$ & $0.00244448$ & $-0.00544448 $\\
$u^2$&&$ 0.107422$ & $0.142112$ & $-0.00962904$ & $0.00269565 $\\
$u^3$&&$ -0.0363689$ & $-0.0545066$ & $0.00975125$ & $-9.31427 \times 10^{-4} $\\
$u^4$&&$ 0.00772009$ & $0.0145342$ & $-0.00442079$ & $0.00173662 $\\
$u^5$&&$ -0.00126492$ & $-0.00291698$ & $0.00128249$ & $-7.61018 \times 10^{-4} $\\
$u^6$&&$ 1.60790 \times 10^{-4}$ & $4.11474 \times 10^{-4}$ & $-2.40437 \times 10^{-4}$ & $1.54324 \times 10^{-4} $\\
$u^7$&&$ -1.38523 \times 10^{-5}$ & $-3.69248 \times 10^{-5}$ & $2.69633 \times 10^{-5}$ & $-1.70449 \times 10^{-5} $\\
$u^8$&&$ 6.87170 \times 10^{-7}$ & $1.86585 \times 10^{-6}$ & $-1.61695 \times 10^{-6}$ & $9.99396 \times 10^{-7} $\\
$u^9$&&$ -1.46371 \times 10^{-8}$ & $-4.02619 \times 10^{-8}$ & $4.00602 \times 10^{-8}$ & $-2.37364 \times 10^{-8} $\\
\hline
\hline
\multicolumn{6}{c}{$^{27}$Al} \\
\multicolumn{6}{c}{$u=p^2b^2/2 \,, \: b=1.8405 \, {\rm fm}$} \\
\hline 
$e^{-u}\times$ & $S_{00}$ & $S_{11}$ (1b+2b min) & $S_{11}$ (1b+2b max) &
$S_{01}$ (1b+2b min) & $S_{01}$ (1b+2b max) \\
\hline
$1$ & $0.0888149$ & $0.0256387$ & $0.0412381$ & $0.0949175$ & $0.121145 $\\
$u$ & $-0.117822$ & $-0.0539361$ & $-0.0881079$ & $-0.152223$ & $-0.212484 $\\
$u^2$ & $0.0631336$ & $0.0638570$ & $0.0973265$ & $0.108925$ & $0.189337 $\\
$u^3$ & $-0.00919554$ & $-0.0473962$ & $-0.0555104$ & $-0.0348055$ & $-0.0898511 $\\
$u^4$ & $5.84421 \times 10^{-4}$ & $0.0242338$ & $0.0200475$ & $0.00826932$ & $0.0309681 $\\
$u^5$ & $5.54484 \times 10^{-4}$ & $-0.00781004$ & $-0.00447580$ & $-0.00135106$ & $-0.00679460 $\\
$u^6$ & $-1.15453 \times 10^{-4}$ & $0.00153205$ & $6.45927 \times 10^{-4}$ & $1.93042 \times 10^{-4}$ & $0.00101787 $\\
$u^7$ & $1.40388 \times 10^{-5}$ & $-1.76118 \times 10^{-4}$ & $-5.82323 \times 10^{-5}$ & $-2.20321 \times 10^{-5}$ & $-9.71893 \times 10^{-5} $\\
$u^8$ & $-9.21830 \times 10^{-7}$ & $1.08574 \times 10^{-5}$ & $3.00602 \times 10^{-6}$ & $1.39046 \times 10^{-6}$ & $5.18194 \times 10^{-6} $\\
$u^9$ & $2.52336 \times 10^{-8}$ & $-2.75875 \times 10^{-7}$ & $-6.68767 \times 10^{-8}$ & $-3.63020 \times 10^{-8}$ & $-1.17607 \times 10^{-7} $\\
\hline
$e^{-u}\times$ & & $S_{p}$ (1b+2b min) & $S_{p}$ (1b+2b max)& 
$S_{n}$ (1b+2b min) & $S_{n}$ (1b+2b max) \\
\hline
$1$&&$ 0.209087$ & $0.251154$ & $0.00893959$ & $0.0192751 $\\
$u$&&$ -0.317485$ & $-0.417183$ & $0.00590871$ & $-0.0132327 $\\
$u^2$&&$ 0.213007$ & $0.344662$ & $-0.0270773$ & $-0.00545593 $\\
$u^3$&&$ -0.0610495$ & $-0.147056$ & $0.0233435$ & $0.0111533 $\\
$u^4$&&$ 0.0133827$ & $0.0465970$ & $-0.00948779$ & $-0.00603345 $\\
$u^5$&&$ -0.00157210$ & $-0.00899070$ & $0.00262032$ & $0.00259059 $\\
$u^6$&&$ 1.66098 \times 10^{-4}$ & $0.00121558$ & $-4.42643 \times 10^{-4}$ & $-6.07533 \times 10^{-4} $\\
$u^7$&&$ -1.51579 \times 10^{-5}$ & $-1.05060 \times 10^{-4}$ & $4.79465 \times 10^{-5}$ & $8.40120 \times 10^{-5} $\\
$u^8$&&$ 8.74763 \times 10^{-7}$ & $5.13463 \times 10^{-6}$ & $-2.80932 \times 10^{-6}$ & $-5.86214 \times 10^{-6} $\\
$u^9$&&$ -2.15130 \times 10^{-8}$ & $-1.07015 \times 10^{-7}$ & $6.92513 \times 10^{-8}$ & $1.64380 \times 10^{-7} $\\
\hline
\end{tabular*}
\end{center}
\end{table}

%\clearpage

\newpage

\onecolumngrid

\begin{table}[t]
\caption{Fits to the isoscalar/isovector structure factors $S_{00}$, 
$S_{11}$ and $S_{01}$ as well as ``proton-only" and ``neutron-only"
structure factors $S_{p}$ and $S_{n}$ for spin-dependent WIMP elastic
scattering off $^{29}$Si,
including 1b and 2b currents as in Fig.~13. The upper and lower limits
from the theoretical error band were used for the fit.  The fitting
function of the dimensionless variable $u = p^2 b^2/2$ is $S_{ij}(u) =
e^{-u} \sum_{n=0}^9 c_{ij,n} u^n$. The rows give the coefficients
$c_{ij,n}$ of the $u^n$ terms in the polynomial.\label{si_fit}}
\begin{center}
\begin{tabular*}{0.773\textwidth}{c||c|c|c|c|c}
\hline
\multicolumn{6}{c}{$^{29}$Si} \\
\multicolumn{6}{c}{$u=p^2b^2/2 \,, \: b=1.8575 \, {\rm fm}$} \\
\hline
$e^{-u}\times$ & $S_{00}$ & $S_{11}$ (1b+2b min) & $S_{11}$ (1b+2b max) & 
$S_{01}$ (1b+2b min) & $S_{01}$ (1b+2b max) \\
\hline
$1$ & $0.0140647$ & $0.00434396$ & $0.00692435$ & $-0.0197473$ & $-0.0155117 $\\
$u$ & $-0.0188522$ & $-0.00978508$ & $-0.0145952$ & $0.0343683$ & $0.0258450 $\\
$u^2$ & $0.0149891$ & $0.0141312$ & $0.0170700$ & $-0.0349170$ & $-0.0268086 $\\
$u^3$ & $-0.00542122$ & $-0.0120045$ & $-0.0101378$ & $0.0178060$ & $0.0173458 $\\
$u^4$ & $0.00117173$ & $0.00602619$ & $0.00368687$ & $-0.00551301$ & $-0.00805050 $\\
$u^5$ & $-1.15932 \times 10^{-4}$ & $-0.00177394$ & $-7.87789 \times 10^{-4}$ & $8.86605 \times 10^{-4}$ & $0.00251057 $\\
$u^6$ & $2.47182 \times 10^{-5}$ & $3.11634 \times 10^{-4}$ & $1.05603 \times 10^{-4}$ & $-7.60246 \times 10^{-5}$ & $-5.25166 \times 10^{-4} $\\
$u^7$ & $-3.04480 \times 10^{-6}$ & $-3.20168 \times 10^{-5}$ & $-8.92530 \times 10^{-6}$ & $1.58691 \times 10^{-6}$ & $6.63557 \times 10^{-5} $\\
$u^8$ & $2.00549 \times 10^{-7}$ & $1.76286 \times 10^{-6}$ & $4.47332 \times 10^{-7}$ & $2.54524 \times 10^{-7}$ & $-4.41639 \times 10^{-6} $\\
$u^9$ & $-5.46011 \times 10^{-9}$ & $-3.97506 \times 10^{-8}$ & $-9.82815 \times 10^{-9}$ & $-1.38615 \times 10^{-8}$ & $1.19592 \times 10^{-7} $\\
\hline
$e^{-u}\times$ & & $S_{p}$ (1b+2b min) & $S_{p}$ (1b+2b max)& 
$S_{n}$ (1b+2b min) & $S_{n}$ (1b+2b max) \\
\hline
$1$&&$ 0.00125408$ & $0.00296249$ & $0.0337976$ & $0.0409244 $\\
$u$&&$ 6.68801 \times 10^{-4}$ & $-0.00455830$ & $-0.0515755$ & $-0.0717867 $\\
$u^2$&&$ -0.00210934$ & $0.00942858$ & $0.0452607$ & $0.0799249 $\\
$u^3$&&$ 0.00149251$ & $-0.0105616$ & $-0.0201013$ & $-0.0491256 $\\
$u^4$&&$ -3.59430 \times 10^{-4}$ & $0.00655559$ & $0.00538148$ & $0.0197508 $\\
$u^5$&&$ -4.73546 \times 10^{-5}$ & $-0.00221187$ & $-7.60569 \times 10^{-4}$ & $-0.00486760 $\\
$u^6$&&$ 4.81182 \times 10^{-5}$ & $4.27089 \times 10^{-4}$ & $9.20786 \times 10^{-5}$ & $7.91536 \times 10^{-4} $\\
$u^7$&&$ -9.10073 \times 10^{-6}$ & $-4.47737 \times 10^{-5}$ & $-9.01250 \times 10^{-6}$ & $-7.77097 \times 10^{-5} $\\
$u^8$&&$ 8.45631 \times 10^{-7}$ & $2.52194 \times 10^{-6}$ & $5.41575 \times 10^{-7}$ & $4.13236 \times 10^{-6} $\\
$u^9$&&$ -3.00417 \times 10^{-8}$ & $-5.89991 \times 10^{-8}$ & $-1.39919 \times 10^{-8}$ & $-9.10412 \times 10^{-8} $\\
\hline

\end{tabular*}
\end{center}
\end{table}
\end{widetext}

\end{document}